%% file: main.tex
\documentclass{article}
\input{preamble}

\begin{document}

\maketitle
\vspace{-1cm}
\section*{Abstract}
Difference-in-differences (DiD) identification relies mainly on a parallel trends assumption about untreated potential outcomes. Researchers often relax this assumption by assuming conditional parallel trends within units with the same covariate values. However, the process of selecting which covariates to include in this assumption is often \emph{ad hoc}. We propose a formal approach to select the variables that support conditional parallel trends based on graphical criteria. We show that the parallel trends assumption is rarely justified without conditioning on covariates, and that unconditional and conditional parallel trends can conflict with one another. We also demonstrate that a time-invariant covariate with a time-invariant effect on the outcome, which might not ordinarily be considered a confounder in DiD, may be a useful conditioning variable. We clarify that adjustment for a post-treatment covariate depends on what causes that covariate to change. Extending our framework to multiple time periods, we distinguish between treatment type and rollout strategy and examine the problem of treatment-confounder feedback. On the estimation side, we argue that the difficulty of incorporating covariates in DiD, often framed as an estimator problem, is more accurately understood as a misalignment between the adjustment set used by the estimator and the adjustment set required for identification. This misalignment affects several popular estimation procedures, and resolving it requires not a change of estimator, but a change in how covariates enter the estimation procedure. We show how to achieve this alignment for all estimators we evaluate.

\clearpage

\section{Introduction}

\ac{DiD} is a popular identification strategy that relies mainly on an assumption that the outcome evolution of the untreated group can stand in for the counterfactual outcome evolution of the treated group had it not been treated. 
This assumption is commonly described as ``parallel trends" in untreated potential outcomes. Although other assumptions are needed for \ac{DiD} identification, including no interference, consistency, and no anticipation \cite{lechnerEstimationCausalEffects2011,rothWhatsTrendingDifferenceindifferences2023}, the parallel trends assumption receives particular scrutiny. For instance, researchers frequently assess its plausibility by testing whether outcome trends are parallel across treated and control groups prior to treatment \cite{bilinski_nothing_2026}. 

Alternatively, we may assume parallel trends holds only for units with the same covariate values, i.e., \emph{conditional} parallel trends. A natural question then arises: how should one select the variables on which to condition? In practice, variables are reported with little justification and the selection often includes every measured variable, i.e, a ``kitchen sink" approach. In studies that give more attention to variable selection, variables may be assessed in isolation, considering whether each evolves differently across groups or has a time-varying relationship to the outcome \cite{abadieSemiparametricDifferenceindifferencesEstimators2005,zeldowConfoundingRegressionAdjustment2021}. 

These approaches resemble historical practice in cross-sectional studies that assume conditional exchangeability (also known as no unmeasured confounding). In this context, each variable was evaluated as a potential confounder by asking whether it was related to both treatment assignment and the outcome. 
More recent work has refined the process, for instance, using \acp{DAG} and Pearl's backdoor criterion \cite{pearlCausalityModelsReasoning2000,pearlCausalInferenceStatistics2016} to select a minimally sufficient set of variables that blocks all backdoor (non-causal) paths between the treatment and outcome \cite{hernanCausalKnowledgePrerequisite2002,tennantUseDirectedAcyclic2021,cinelliCrashCourseGood2024a}.
Such a set avoids the pitfalls of too many control variables, such as limited positivity \cite{damourOverlapObservationalStudies2021}, collider bias \cite{cinelliCrashCourseGood2024a} and M-bias \cite{dingAdjustNotAdjust2015}.

In this work, we extend these ideas to \ac{DiD} by proposing a graphical approach to selecting variables that support conditional parallel trends. We examine the relationship between unconditional and conditional parallel trends, consider the role of time-invariant and time-varying covariates in identification, including post-treatment covariate values, and extend the framework to settings with multiple time periods. We also evaluate how several popular \ac{DiD} estimators handle adjustment variables and propose ways to better align estimation with identification.

The remainder of this article is organized as follows. Section \ref{SecRL} reviews the related literature and situates our specific contributions within it. Section \ref{SecPTECB} establishes the equivalence between parallel trends and equi-confounding, which enables us to represent this key identifying assumption in a causal diagram in Section \ref{SecDiDgraph}, where we show \ac{DiD} identification using graphical criteria. In Section \ref{SecCondlPT}, we extend the framework to scenarios with covariates and multiple time periods. Section \ref{SecEst} compares popular estimation methods and examines how they handle covariates, including the modifications needed to achieve alignment with identification requirements. Section \ref{SecRes} presents the results of a simulation study, and Section \ref{SecDis} concludes with a discussion.

\section{Related literature}\label{SecRL}

We build on recent work that uses causal diagrams for \ac{DiD} identification.
Chalak introduced the concept of ``equi-confounding", in which the same unobserved confounder affects equally (e.g., proportionally) one or more observed variables, including covariates and outcome variables \cite{chalakIdentificationExogeneityEquiconfounding2013}.
Zhang et al.~later noted that \ac{DiD} is a special case of equi-confounding where two outcome variables (i.e., the pre- and post-treatment) are equally affected by the unobserved confounder \cite{zhangExploitingEqualityConstraints2021}. Using the same assumption of equi-confounding, Sofer et al.~showed that \ac{DiD} can be viewed as a negative outcome control strategy in which the pre-treatment outcome is the negative control affected by the same confounding as the post-treatment outcome \cite{soferNegativeOutcomeControl2016a}. Those authors distinguished between additive equi-confounding and two alternatives, quantile-quantile and distributional equi-confounding, which are useful when pre- and post-treatment outcomes are measured on different scales. Since this situation is not typical in \ac{DiD}, we assume additivity and omit this qualifier from the term equi-confounding hereafter.

Without using the term equi-confounding, other authors have independently noted the need for the same unobserved confounding to affect pre- and post-treatment outcome variables in \ac{DiD} \cite{weberAssumptionTradeOffsWhen2015a,kimGainScoresRevisited2021a,rensonUsingCausalDiagrams2025}. Weber et al.~highlighted this requirement while discussing the trade-offs between parallel trends and exchangeability \cite{weberAssumptionTradeOffsWhen2015a}. Kim and Steiner showed that \ac{DiD} identification relies on ``offsetting rather than blocking" backdoor paths, such that equality of the confounding in the pre- and post-treatment periods allows the bias to cancel \cite{kimGainScoresRevisited2021a}. Renson et al.~formalized the structural conditions for unconditional parallel trends using \acp{SWIG}, further clarifying the graphical basis of \ac{DiD} identification \cite{rensonUsingCausalDiagrams2025}.

A related strand of work has considered conditioning on covariates in \ac{DiD}. One of the earliest estimation procedures designed to handle time-varying confounding in \ac{DiD} was proposed by Stuart et al., who combined propensity score weighting with a subsequent \ac{TWFE} weighted regression. More recently, Ghanem et al.~explored mechanisms of selection into treatment under which unconditional or conditional parallel trends would hold and noted that ``incorporating time-varying covariates makes the restrictions on the selection mechanism more plausible" \cite{ghanemSelectionParallelTrends2025}. Caetano and Callaway demonstrated the limitations of the standard \ac{TWFE} model for incorporating covariates in \ac{DiD} when both time-invariant and time-varying covariates need to be adjusted for, and proposed alternative estimators paired with adjustment sets containing baseline covariates plus either the post-treatment value or the change in the covariate \cite{caetanoDifferenceinDifferencesWhenParallel2024}. Myint examined the relationship between identification under sequential exchangeability and parallel trends, and proposed hybrid estimators that use sequential \ac{IPW} within a \ac{DiD} framework \cite{myintControllingTimevaryingConfounding2023}. Renson et al.~further formalized the connection between g-methods and \ac{DiD} by deriving an identification solution and appropriate estimators for sustained treatments under parallel trends \cite{rensonIdentifyingEstimatingEffects2023}. In related work, Shahn et al. extended structural nested models to \ac{DiD} under the parallel trends assumption \cite{shahnStructuralNestedMean2022}.

In this work, we aim to provide actionable guidance for variable selection in \ac{DiD} and align estimation with identification.
Thus, our first contribution is to provide a formal variable selection framework that makes conditional parallel trends plausible. 
We extend the graph-based \ac{DiD} literature to more complex settings than previously considered, including multiple covariates and  time periods. 
Doing so allows us to derive several new insights. 
First, we show how strong unconditional parallel trends is and highlight a potential conflict between unconditional and conditional parallel trends. 
Second, we demonstrate that a time-invariant covariate with a time-invariant effect on the outcome can serve as a valid adjustment variable for time-varying confounding in \ac{DiD}. 
Third, we clarify that the appropriateness of adjusting for post-treatment covariate values, often excluded by default in most \ac{DiD} studies, depends on the causes of the covariate change. 
Fourth, when extending our models to multiple time periods, we distinguish between treatment type and rollout strategy, and examine the problem of treatment-confounder feedback, as well as strategies for avoiding it.

Our second key contribution concerns estimation and its alignment with identification. A recurring pattern in the \ac{DiD} literature attributes the difficulty of accommodating covariates to limitations of particular estimators. This concern arises most often with time-varying covariates, though for certain estimators it also extends to time-invariant covariates. We show that the underlying problem is more accurately understood as a misalignment between the adjustment set used by the estimator and the adjustment set required for identification. This misalignment affects several popular estimation procedures, including but not limited to the standard \ac{TWFE} specification. The implication is that handling covariates in \ac{DiD} is less a matter of estimator choice than of ensuring that each estimator operates on the adjustment set required for identification. For every estimator we consider, we demonstrate how this alignment can be achieved, recovering valid estimates without modifying the estimator's underlying structure.

\section{Parallel trends and equi-confounding}\label{SecPTECB}

First, we introduce notation. Let $i \in \{1, \dots, N\}$ index units (e.g., individuals or regions), and $t \in \{0, 1\}$ index time, where $t = 0$ is before treatment and $t = 1$ is after. Each unit is observed at both time points (or, if we have cross-sectional data, each cluster or unit of treatment assignment is observed at both time points). Let $A_{i1} \in \{0,1\}$ represent a static treatment first introduced at $t=1$, where $A_{i1}=1$ if unit $i$ receives treatment at $t=1$ and $A_{i1}=0$ if it does not (i.e., is a ``control'' unit). We use $Y_{it}$ to denote the observed outcome for unit $i$ at time $t$ and $Y_{it}(0), Y_{it}(1)$, the potential outcomes under control and treatment, respectively. We assume that treatment occurs before the outcome observed at $t=1$. The causal estimand of interest is the \ac{ATT},
\begin{equation}\label{eq:att}
\mathbb{E}[Y_{i1}(1) - Y_{i1}(0) | A_{i1} = 1].
\end{equation}

To identify this quantity, we first make three preliminary assumptions \cite{lechnerEstimationCausalEffects2011,rothWhatsTrendingDifferenceindifferences2023}.
First, we assume that future treatments do not affect past outcomes (no anticipation),
so that the observed pre-treatment outcomes are untreated potential outcomes for all units,
\begin{equation}\label{eq:att2}
Y_{i0} = Y_{i0}(0) \; \forall \; i.
\end{equation}

Second, we assume the treatment assignment of one unit does not affect the potential outcomes of any other unit (no interference). Third, we assume that there are no hidden levels of treatment. This assumption allows us to substitute observed outcomes for the potential outcomes given the corresponding observed treatment levels (consistency),
\begin{equation}\label{eq:att3}
Y_{it} = A_{it} Y_{it}(1) + (1-A_{it}) Y_{it}(0).
\end{equation}

Hereafter, we will omit the unit subscript $i$ to simplify notation, writing $Y_t(a)$ and $A_t$ in place of $Y_{it}(a)$ and $A_{it}$, respectively.

Next, we assume parallel trends in untreated potential outcomes between the treated and control groups,
\begin{equation}\label{eq:pt}
\mathbb{E}[Y_1(0)-Y_0(0)|A_1=1] = \mathbb{E}[Y_1(0)-Y_0(0)|A_1=0].
\end{equation}

A useful reframing of the parallel trends assumption is in terms of equi-confounding, i.e., assuming that the difference in untreated potential outcomes between treated and control groups is equal at $t=1$ and $t=0$,
\begin{flalign}\label{eq:ecb}
\mathbb{E}[Y_1(0)|A_1=1]-\mathbb{E}[Y_1(0)|A_1=0] & = \mathbb{E}[Y_0(0)|A_1=1]-\mathbb{E}[Y_0(0)|A_1=0].
\end{flalign}

Using either Eq.~(\ref{eq:pt}) or (\ref{eq:ecb}) plus the additional assumptions above, we solve for the counterfactual,
\begin{flalign*}
\mathbb{E}[Y_1(0) | A_1 = 1] & = \mathbb{E}[Y_0 (0) | A_1 = 1] + \mathbb{E}[Y_1 (0) - Y_0 (0) | A_1 = 0]\\
	          		& = \mathbb{E}[Y_0 (1) | A_1 = 1] + \mathbb{E}[Y_1 (0) - Y_0 (0) | A_1 = 0]\\
  & = \mathbb{E}(Y_0 | A_1 = 1) + \mathbb{E}(Y_1 - Y_0 | A_1 = 0),
\end{flalign*}
and plug it into the causal estimand in Eq.~(\ref{eq:att}) to obtain the statistical estimand,
\begin{equation}\label{eq:sestpt}
\mathbb{E}(Y_1 - Y_0 | A_1 = 1) - \mathbb{E}(Y_1 - Y_0 | A_1 = 0).
\end{equation}

\section{Graph-based DiD identification}\label{SecDiDgraph}

\begin{figure}[ht]
\centering
\begin{tikzpicture}[>=triangle 45, font=\scriptsize]
\node[fill,circle,inner sep=0pt,minimum size=5pt,label={left:{$A_1$}}] (A1) at (0,0) {};
\node[fill,circle,inner sep=0pt,minimum size=5pt,label={left:{$Y_0$}}] (Y0) at (-2,0) {};
\node[fill,circle,inner sep=0pt,minimum size=5pt,label={right:{$Y_1$}}] (Y1) at (2,0) {};
\node[draw,circle,inner sep=0pt,minimum size=5pt,label={left:{$V_0$}}] (V0) at (-2.25,1) {};
\node[draw,circle,inner sep=0pt,minimum size=5pt,label={left:{$S_0$}}] (S0) at (-2.2,1.7) {};
\node[fill,circle,inner sep=0pt,minimum size=5pt,label={right:{$\Delta Y$}}] (DELTAY) at (2,-1) {};
\node[minimum size=5pt,label={left:{$*$}}] (*) at (0.25,1) {};
\node[minimum size=5pt,label={left:{$*$}}] (*) at (-2,0.6) {};
\node[minimum size=5pt,label={left:{$1$}}] (1) at (2.5,-0.4) {};
\node[minimum size=5pt,label={left:{$-1$}}] (-1) at (-0.6,-0.85) {};
\draw[->,shorten >= 1pt] (A1)--(Y1);
\draw[->,shorten >= 1pt] (V0)--(Y0);
\draw[-{Latex[length=6.5pt]}] (Y0) to[bend right=19] (DELTAY);
\draw[->,shorten >= 1pt] (Y1)--(DELTAY);
\draw[-{Latex[length=6.5pt]}] (V0) to[bend left=20] (A1);
\draw[-{Latex[length=6.5pt]}] (V0) to[bend left=15] (Y1);
\draw[->,shorten >= 1pt] (S0)--(Y0);
\draw[-{Latex[length=6.5pt]}] (S0) to[bend left=15] (Y1);
\end{tikzpicture}
\caption{Causal diagram in which equi-confounding by the unobserved $V_0$ is denoted by the equal magnitude $*$ on the edges to the outcome at both time periods.}\label{figcdbasic}
\end{figure}
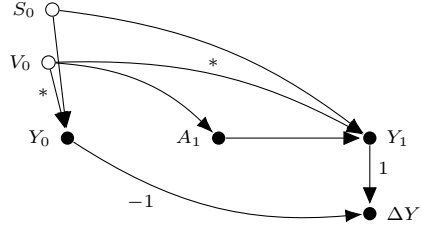

This identification process, solving for the counterfactual using a scale-dependent assumption about evolution of potential outcomes, is standard in the \ac{DiD} literature. 
Next, we consider an alternative route to identification that uses graphical criteria.
Figure \ref{figcdbasic} shows a simple \ac{DAG} relevant for a \ac{DiD} study.
To the observable variables noted above, we add two unobservable variables.
The first unobservable, $V_0$, affects the treatment and outcomes at both times.
We represent the equi-confounding assumption by the equal magnitude $*$ on the edges from $V_0$ to $Y_0$ and $Y_1$.
The second unobservable, $S_0$ affects only outcomes at both times, inducing correlation between them.
We also include the change in the outcome over time, $\Delta Y = Y_1-Y_0$, 
and deterministic coefficients on the edges from $Y_0$ and $Y_1$ to $\Delta Y$. The unmeasured error terms are excluded for simplicity.

To identify the average affect of $A_1$ on $Y_1$, we must block the backdoor path, $A_1 \leftarrow V_0 \xrightarrow{*} Y_1$. However, this path cannot be blocked because $V_0$ is not measured. This challenge motivates a \ac{DiD} identification strategy, in which we leverage pre-treatment outcome measurements at $t=0$ and assume that the confounding bias transmitted through $A_1 \leftarrow V_0 \xrightarrow{*} Y_0$ is of equal magnitude to that operating at $t=1$. Under this equi-confounding assumption, differencing the na\"ive estimates across the two time periods cancels the bias, yielding an unbiased estimate of the treatment effect. Notice that the identification result in Eq.~(\ref{eq:sestpt}) is equivalent to 
\begin{equation}\label{eq:sestpt_alt}
\underset{\text{(treatment effect + confounding)}}{\left[\mathbb{E}(Y_1|A_1=1)-\mathbb{E}(Y_1|A_1=0)\right]}
-
\underset{\text{(confounding)}}{\left[\mathbb{E}(Y_0|A_1=1)-\mathbb{E}(Y_0|A_1=0)\right]}.
\end{equation}

\subsection{Equivalence of estimands}
While we have thus far focused on the effect of $A_1$ on $Y_1$, we can equivalently work with the estimand defined as the effect of $A_1$ on $\Delta Y$. We now establish this equivalence,
\begin{flalign*}
ATT_{\Delta Y} & = \mathbb{E}[\Delta Y(1)-\Delta Y(0) | A_1 = 1]\\
& = \mathbb{E}[Y_1(1)-Y_0(1)-[Y_1(0)-Y_0(0)] | A_1 = 1]\\
& = \mathbb{E}[Y_1(1)-Y_1(0) | A_1 = 1] - \mathbb{E}[Y_0(1)-Y_0(0) | A_1 = 1]\\
& = \mathbb{E}[Y_1(1)-Y_1(0) | A_1 = 1] \hspace{0.2cm} (ATT_{Y_1}),
\end{flalign*}
where $\mathbb{E}[Y_0(1)-Y_0(0)]=0$ under the assumption of no treatment anticipation.

In this context, there are two backdoor paths from treatment to the outcome change,
\begin{flalign}
\label{eq:bdpDeltaY1}
A_1 \leftarrow V_0 \xrightarrow{*} Y_1 \xrightarrow{1} \Delta Y,\\
\label{eq:bdpDeltaY2}
A_1 \leftarrow V_0 \xrightarrow{*} Y_0 \xrightarrow{-1} \Delta Y.
\end{flalign}
Under equi-confounding, these two paths are of equal strength and opposite in sign, making the net confounding zero. 
We can also observe that the statistical estimand in Eq.~(\ref{eq:sestpt}) is equivalent to,
\begin{equation}\label{eq:sestpt_alt_2}
\mathbb{E}(\Delta Y | A_1 = 1) - \mathbb{E}(\Delta Y | A_1 = 0).
\end{equation} 

This graph-based approach to \ac{DiD} identification allows us to see a relatively new point from the \ac{DiD} literature: if the pre-treatment outcome directly affect treatment assignment, parallel trends will not hold \cite{weberAssumptionTradeOffsWhen2015a,imaiWhenShouldWe2019a,kimGainScoresRevisited2021a,rensonUsingCausalDiagrams2025}. To understand this point, note that if $Y_0 \rightarrow A_1$ were present in the graph of Figure \ref{figcdbasic}, we would need to condition on $Y_0$ to block the backdoor path $A_1 \leftarrow Y_0 \rightarrow \Delta Y$. However, by conditioning on $Y_0$, the path in (\ref{eq:bdpDeltaY2}) would also be blocked. Yet, the path in (\ref{eq:bdpDeltaY1}) would remain open and no longer offset. Thus, \ac{DiD} requires that we assume no direct effect of $Y_0$ on $A_1$. This observation helps clarify the controversy over matching on the pre-treatment outcome \cite{chabe-ferretShouldWeCombine2017,dawMatchingRegressionMean2018,ryanWellBalancedTooMatchy2018,hamBenefitsCostsMatching2024}.

\subsection{The compact representation of the causal diagram}

\begin{figure}[ht]
\centering
\begin{tikzpicture}[>=triangle 45, font=\scriptsize]
\node[fill,circle,inner sep=0pt,minimum size=5pt,label={left:{$A_1$}}] (A1) at (0,0) {};
\node[fill,circle,inner sep=0pt,minimum size=5pt,label={right:{$\Delta Y$}}] (DELTAY) at (2,0) {};
\node[draw,circle,inner sep=0pt,minimum size=5pt,label={left:{$V_0$}}] (V0) at (-0.2,1) {};
\node[draw,circle,inner sep=0pt,minimum size=5pt,label={left:{$S_0$}}] (S0) at (1.8,1) {};
\draw[->,shorten >= 1pt] (A1)--(DELTAY);
\draw[->,shorten >= 1pt] (V0)--(A1);
\draw[->,shorten >= 1pt] (S0)--(DELTAY);
\end{tikzpicture}
\caption{Causal diagram in which equi-confounding is represented by the absence of an arrow from the unmeasured confounder, $V_0$, to the change in the outcome over time, $\Delta Y$.}\label{figcdbasic_alt}
\end{figure}
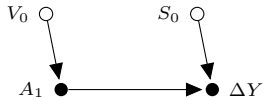

We can also construct a compact representation of the natural causal diagram from Figure \ref{figcdbasic} by omitting $Y_0$ and $Y_1$ and connecting the remaining nodes directly to $\Delta Y$, as shown in Figure \ref{figcdbasic_alt}.

In this representation, equi-confounding is encoded as a single missing edge from $V_0$ to $\Delta Y$, rather than an equality constraint across two backdoor paths. By inspection, there are no backdoor paths from $A_1$ to $\Delta Y$. As we will show in Section \ref{SecCondlPT}, this compact representation becomes particularly valuable under conditional parallel trends, where the backdoor criterion guides the selection of a sufficient adjustment set.

\subsection{Algebraic verification using a linear \ac{SCM}}
Thus far, we have not imposed any functional form restrictions. We now consider a linear \ac{SCM} to express the identification result algebraically in terms of structural coefficients. Under a linear \ac{SCM}, we can assign a structural coefficient to each edge of the graph and apply Wright's path tracing rules to express the covariance between any two variables as the sum of products of structural coefficients traced across paths connecting them \cite{wrightCorrelationCausation1921,chenGraphicalToolsLinear2018,cinelliCrashCourseGood2024a}. In a linear model with a single regressor, the regression coefficient is $\beta = \sigma(Y_1,A_1)/\text{Var}(A_1)$. For expositional clarity, we assume unit variances throughout, so that covariances correspond directly to regression coefficients.

\begin{figure}[ht]
\begin{subfigure}{.4\textwidth}
\hspace{2cm}
\begin{tikzpicture}[>=triangle 45, font=\scriptsize]
\node[fill,circle,inner sep=0pt,minimum size=5pt,label={left:{$A_1$}}] (A1) at (0,0) {};
\node[fill,circle,inner sep=0pt,minimum size=5pt,label={left:{$Y_0$}}] (Y0) at (-2,0) {};
\node[fill,circle,inner sep=0pt,minimum size=5pt,label={right:{$Y_1$}}] (Y1) at (2,0) {};
\node[draw,circle,inner sep=0pt,minimum size=5pt,label={left:{$V_0$}}] (V0) at (-2.25,1) {};
\node[draw,circle,inner sep=0pt,minimum size=5pt,label={left:{$S_0$}}] (S0) at (-2.2,1.7) {};
\node[fill,circle,inner sep=0pt,minimum size=5pt,label={right:{$\Delta Y$}}] (DELTAY) at (2,-1) {};
\node[minimum size=5pt,label={left:{$e$}}] (e) at (0.25,1.4) {};
\node[minimum size=5pt,label={left:{$c$}}] (c) at (0.25,1) {};
\node[minimum size=5pt,label={left:{$c$}}] (c) at (-2,0.6) {};
\node[minimum size=5pt,label={left:{$a$}}] (a) at (1.15,0.15) {};
\node[minimum size=5pt,label={left:{$d$}}] (d) at (-2,1.35) {};
\node[minimum size=5pt,label={left:{$b$}}] (b) at (-0.7,0.55) {};
\node[minimum size=5pt,label={left:{$1$}}] (1) at (2.5,-0.4) {};
\node[minimum size=5pt,label={left:{$-1$}}] (-1) at (-0.6,-0.85) {};
\draw[->,shorten >= 1pt] (A1)--(Y1);
\draw[->,shorten >= 1pt] (V0)--(Y0);
\draw[-{Latex[length=6.5pt]}] (Y0) to[bend right=19] (DELTAY);
\draw[->,shorten >= 1pt] (Y1)--(DELTAY);
\draw[-{Latex[length=6.5pt]}] (V0) to[bend left=20] (A1);
\draw[-{Latex[length=6.5pt]}] (V0) to[bend left=15] (Y1);
\draw[->,shorten >= 1pt] (S0)--(Y0);
\draw[-{Latex[length=6.5pt]}] (S0) to[bend left=15] (Y1);
\end{tikzpicture}
\caption{ }
\end{subfigure}
\hspace{2cm}
\begin{subfigure}{.4\textwidth}
\centering
\begin{tikzpicture}[>=triangle 45, font=\scriptsize]
\node[fill,circle,inner sep=0pt,minimum size=5pt,label={left:{$A_1$}}] (A1) at (0,0) {};
\node[fill,circle,inner sep=0pt,minimum size=5pt,label={right:{$\Delta Y$}}] (DELTAY) at (2,0) {};
\node[draw,circle,inner sep=0pt,minimum size=5pt,label={left:{$V_0$}}] (V0) at (-0.2,1) {};
\node[draw,circle,inner sep=0pt,minimum size=5pt,label={left:{$S_0$}}] (S0) at (1.8,1) {};
\node[minimum size=5pt,label={left:{$a$}}] (a) at (1.25,0.15) {};
\node[minimum size=5pt,label={left:{$e-d$}}] (e-d) at (2.85,0.6) {};
\node[minimum size=5pt,label={left:{$b$}}] (b) at (0,0.6) {};
\draw[->,shorten >= 1pt] (A1)--(DELTAY);
\draw[->,shorten >= 1pt] (V0)--(A1);
\draw[->,shorten >= 1pt] (S0)--(DELTAY);
\end{tikzpicture}
\vspace{0.5cm}
\caption{ }
\end{subfigure}
\caption{Same model as Figure \ref{figcdbasic} but now assuming a linear structural causal model, indicated by coefficients on each edge. \textbf{(a)} Natural representation. \textbf{(b)} Compact representation.}
\label{figcdbasic2_figcdbasic_alt_linear}
\end{figure}
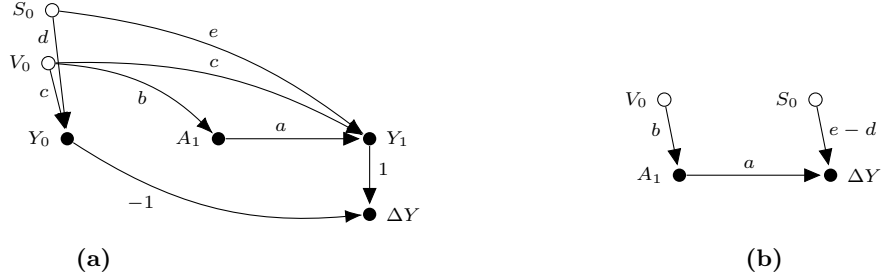

Figure \ref{figcdbasic2_figcdbasic_alt_linear} revisits the causal diagrams from Figures \ref{figcdbasic} and \ref{figcdbasic_alt}, now with structural coefficients assigned to each edge under a linear SCM. The magnitude of the structural coefficients in the compact representation (Panel b) follow from the covariances implied by the natural causal diagram (Panel a). Under equi-confounding, $\sigma(\Delta Y,V_0)= c \cdot (1) + c \cdot (-1) = 0$, confirming the absence of the edge from $V_0$ to $\Delta Y$. The remaining edges carry magnitudes $\sigma(A_1,V_0)= b$ and $\sigma(\Delta Y, S_0)=e-d$.

Applying Wright's rules to the natural diagram in Panel (a), we first consider the effect of $A_1$ on $Y_1$. The covariance $\sigma(Y_1, A_1)=a+bc$ includes the treatment effect $a$ and the confounding bias $bc$, while $\sigma(Y_0, A_1)=bc$ captures confounding alone. Their difference recovers the treatment effect $a$. We can equivalently consider the effect of $A_1$ on $\Delta Y$ directly, $\sigma(\Delta Y, A_1) = a + bc - bc = a$, which simplifies to the treatment effect. Turning to the compact representation in Panel (b), Wright's rules yield the same result: $\sigma(\Delta Y, A_1) = a$, since there are no backdoor paths between $A_1$ and $\Delta Y$, the treatment effect is immediately identified.

\subsection{Summary}
To summarize, we have described \ac{DiD} identification through several complementary lenses. First, using potential outcomes, we proceeded from the parallel trends assumption to obtain the statistical estimand. Second, we established the equivalence between parallel trends and equi-confounding, allowing us to encode this assumption in a causal diagram and derive the identification result graphically. In the natural representation of the causal diagram, the effect of $A_1$ on $Y_1$ captures both treatment effect and confounding, while the effect of $A_1$ on $Y_0$ captures confounding alone, so that their difference recovers the treatment effect. Focusing on the equivalent estimand, the effect of $A_1$ on $\Delta Y$, reveals two backdoor paths of equal strength and opposite sign that cancel under equi-confounding. This observation motivated the compact representation, in which $Y_0$ an $Y_1$ are omitted and equi-confounding is encoded by the absence of an edge from $V_0$ to $\Delta Y$, leaving no open backdoor paths from $A_1$ to $\Delta Y$. Third, we introduced a linear \ac{SCM} and Wright's rules to retrace these same steps algebraically, assigning structural coefficients to each edge and obtaining the identification result through covariance calculations on both representations. The linear \ac{SCM} serves a purely expositional role. It allows us to decompose complex settings and inspect underlying mechanisms in later sections, but the identification result itself does not depend on functional form assumptions. Having established that the effect of $A_1$ on $\Delta Y$ is equivalent to the effect of $A_1$ on $Y_1$, we henceforth work with $\Delta Y$ as the outcome.

\section{Conditional parallel trends or conditional equi-confounding}\label{SecCondlPT}

The parallel trends assumption in Eq.~(\ref{eq:pt}) can be relaxed by assuming it holds within strata defined by $X$,
\begin{equation}
\mathbb{E}[Y_1(0)-Y_0(0)|A_1=1,X] = \mathbb{E}[Y_1(0)-Y_0(0)|A_1=0,X],
\end{equation}
where $X$ is a vector that may include both time-invariant and time-varying covariates.
Equivalently, equi-confounding in Eq.~(\ref{eq:ecb}) can be relaxed by conditioning on covariates,
\begin{equation}
\mathbb{E}[Y_1(0)|A_1=1,X]-\mathbb{E}[Y_1(0)|A_1=0,X] = \mathbb{E}[Y_0(0)|A_1=1,X]-\mathbb{E}[Y_0(0)|A_1=0,X].
\end{equation}

Using these conditional assumptions for causal identification requires us to also assume positivity (also known as overlap or common support), so that for each value of $X$ there is a positive probability of receiving each treatment level. 
With the other assumptions introduced earlier (no anticipation, no interference, and consistency) and the same algebraic steps, we can identify the counterfactual outcome and obtain the statistical estimand,
\begin{equation}
\label{eq:attX1}
\mathbb{E_{X}}[\mathbb{E}(Y_1 - Y_0 | A_1 = 1, X) - \mathbb{E}(Y_1 - Y_0 | A_1 = 0, X)].
\end{equation}

The question is then how to select the set of covariates $X$ that makes this conditional parallel trends (or equi-confounding) assumption plausible. The compact representation of the causal diagram is particularly well suited to this task. Since it encodes equi-confounding directly through the absence of an edge, rather than equality constraints, the standard backdoor criterion can be applied to determine sufficient adjustment sets. Importantly, this graphical approach to variable selection is fully general, i.e., it does not require parametric restrictions on the structural equations. The structural coefficients displayed on the edges of our causal diagrams throughout the manuscript are intended solely to make the underlying mechanics transparent.

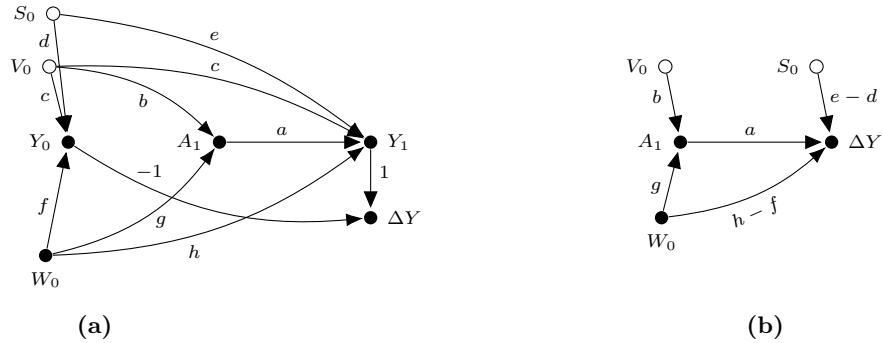
\begin{figure}[ht]
\begin{subfigure}{.4\textwidth}
\hspace{2cm}
\begin{tikzpicture}[>=triangle 45, font=\scriptsize]
\node[fill,circle,inner sep=0pt,minimum size=5pt,label={left:{$A_1$}}] (A1) at (0,0) {};
\node[fill,circle,inner sep=0pt,minimum size=5pt,label={left:{$Y_0$}}] (Y0) at (-2,0) {};
\node[fill,circle,inner sep=0pt,minimum size=5pt,label={right:{$Y_1$}}] (Y1) at (2,0) {};
\node[draw,circle,inner sep=0pt,minimum size=5pt,label={left:{$V_0$}}] (V0) at (-2.25,1) {};
\node[draw,circle,inner sep=0pt,minimum size=5pt,label={left:{$S_0$}}] (S0) at (-2.2,1.7) {};
\node[fill,circle,inner sep=0pt,minimum size=5pt,label={right:{$\Delta Y$}}] (DELTAY) at (2,-1) {};
\node[fill,circle,inner sep=0pt,minimum size=5pt,label={below:{$W_0$}}] (W0) at (-2.3,-1.5) {};
\node[minimum size=5pt,label={left:{$e$}}] (e) at (0.25,1.4) {};
\node[minimum size=5pt,label={left:{$c$}}] (c) at (0.25,1) {};
\node[minimum size=5pt,label={left:{$c$}}] (c) at (-2,0.6) {};
\node[minimum size=5pt,label={left:{$a$}}] (a) at (1.15,0.15) {};
\node[minimum size=5pt,label={left:{$d$}}] (d) at (-2,1.35) {};
\node[minimum size=5pt,label={left:{$b$}}] (b) at (-0.7,0.55) {};
\node[minimum size=5pt,label={left:{$1$}}] (1) at (2.5,-0.4) {};
\node[minimum size=5pt,label={left:{$-1$}}] (-1) at (-0.5,-0.4) {};
\node[minimum size=5pt,label={left:{$f$}}] (f) at (-2,-0.85) {};
\node[minimum size=5pt,label={left:{$h$}}] (h) at (0,-1.45) {};
\node[minimum size=5pt,label={left:{$g$}}] (g) at (-0.45,-1.05) {};
\draw[->,shorten >= 1pt] (A1)--(Y1);
\draw[->,shorten >= 1pt] (V0)--(Y0);
\draw[-{Latex[length=6.5pt]}] (Y0) to[bend right=19] (DELTAY);
\draw[->,shorten >= 1pt] (Y1)--(DELTAY);
\draw[-{Latex[length=6.5pt]}] (V0) to[bend left=20] (A1);
\draw[-{Latex[length=6.5pt]}] (V0) to[bend left=15] (Y1);
\draw[->,shorten >= 1pt] (S0)--(Y0);
\draw[-{Latex[length=6.5pt]}] (S0) to[bend left=15] (Y1);
\draw[->,shorten >= 1pt] (W0)--(Y0);
\draw[-{Latex[length=6.5pt]}] (W0) to[bend right=17] (Y1);
\draw[-{Latex[length=6.5pt]}] (W0) to[bend right=20] (A1);
\end{tikzpicture}
\caption{ }
\end{subfigure}
\hspace{2cm}
\begin{subfigure}{.4\textwidth}
\centering
\begin{tikzpicture}[>=triangle 45, font=\scriptsize]
\node[fill,circle,inner sep=0pt,minimum size=5pt,label={left:{$A_1$}}] (A1) at (0,0) {};
\node[fill,circle,inner sep=0pt,minimum size=5pt,label={right:{$\Delta Y$}}] (DELTAY) at (2,0) {};
\node[draw,circle,inner sep=0pt,minimum size=5pt,label={left:{$V_0$}}] (V0) at (-0.2,1) {};
\node[draw,circle,inner sep=0pt,minimum size=5pt,label={left:{$S_0$}}] (S0) at (1.8,1) {};
\node[fill,circle,inner sep=0pt,minimum size=5pt,label={below:{$W_0$}}] (W0) at (-0.25,-1) {};
\node[minimum size=5pt,label={left:{$a$}}] (a) at (1.25,0.15) {};
\node[minimum size=5pt,label={left:{$e-d$}}] (e-d) at (2.85,0.6) {};
\node[minimum size=5pt,label={left:{$b$}}] (b) at (0,0.6) {};
\node[minimum size=5pt,label={left:{$g$}}] (g) at (0,-0.6) {};
\node[minimum size=5pt,label={left, rotate=24:{$h-f$}}] (h-f) at (1.55,-0.75) {};
\draw[->,shorten >= 1pt] (A1)--(DELTAY);
\draw[->,shorten >= 1pt] (V0)--(A1);
\draw[->,shorten >= 1pt] (S0)--(DELTAY);
\draw[-{Latex[length=6.5pt]}] (W0) to[bend right=17] (DELTAY);
\draw[->,shorten >= 1pt] (W0)--(A1);
\end{tikzpicture}
\vspace{0.5cm}
\caption{ }
\end{subfigure}
\caption{Scenario with a baseline covariate $W_0$ having a time-varying relationship to outcomes. \textbf{(a)} Natural representation. \textbf{(b)} Compact representation. Illustrative linear model coefficients are shown as lowercase letters on each edge.}
\label{figcdbasic_alt_linear_cov}
\end{figure}

To illustrate, consider a baseline covariate $W_0$ that has a time-varying effect on the outcome, as shown in Figure \ref{figcdbasic_alt_linear_cov}. Panel (a) includes the natural causal diagram and Panel (b) shows its compact representation.

The arrow from $W_0$ to $\Delta Y$ is retained in the compact representation in Panel (b) because the effect of $W_0$ on the outcome is not constant across time periods. In the linear case, the edge has magnitude $h-f$, ensuring that the covariance $\sigma(\Delta Y, W_0)= ga + h-f$ is preserved across both representations. The presence of this edge creates a backdoor path between $A_1$ and $\Delta Y$ through $W_0$, meaning unconditional parallel trends does not hold. Of note, in the more general (potential nonlinear) case without assumptions about the constancy of effects over time, all edges to $\Delta Y$ in the compact representation would be retained by default.

To restore identification, we turn to conditional parallel trends. Examining the compact representation in Panel (b), we observed that conditioning on $W_0$ blocks all backdoor paths from $A_1$ to $\Delta Y$, and $W_0$ is not a descendant of the treatment. Thus, by the backdoor criterion, \{$W_0$\} is a sufficient adjustment set for identifying the effect of $A_1$ on $\Delta Y$.

As an algebraic complement under linearity, the same result can be obtained by deriving the relevant covariances using Wright's rules and applying Cramér's covariance formula to compute the partial regression coefficient for the effect of $A_1$ on $\Delta Y$, conditional on $W_0$,
\begin{equation*}
\begin{split}
\beta (\Delta Y,A_1|W_0) & = \frac{\sigma(\Delta Y, A_1) - \sigma(\Delta Y,W_0) \sigma(A_1,W_0)}{1-\sigma(A_1,W_0)^2} \\
& = \frac{a+gh-gf - (ga+h-f) (g)}{1-g^2} \\
& = \frac{a+gh-gf -ag^2-gh+gf}{1-g^2} \\
& = \frac{a(1-g^2)}{1-g^2} \\
& = a\;,
\end{split}
\end{equation*}
which recovers the true treatment effect.

In the remainder of this section, we apply both approaches to variable selection across a series of scenarios. Section \ref{ss:time-invariant-only} examines the plausibility of unconditional parallel trends and its potential conflict with conditional parallel trends. Section \ref{ss:time-varying-effect} focuses on the role of time-invariant covariates in addressing time-varying confounding (Section \ref{ss:time-varying-effect}) whereas Section \ref{ss:post-trt-vars} addresses the conditions under which post-treatment covariate values may be adjusted for in a \ac{DiD} study. Finally, Section \ref{ss:mult-time-periods} considers the problem of treatment-confounder feedback in the context of multiple time periods. Of note, we present identification results primarily using graphical criteria applied to the compact representation, particularly from Section \ref{ss:time-varying-effect} onward. Alternative proofs for each scenario using Wright's rules and Cramér's covariance formula are provided in the Appendix.

\subsection{Unconditional vs. conditional parallel trends}\label{ss:time-invariant-only}

\begin{figure}[h!]
\begin{subfigure}{.4\textwidth}
\hspace{2cm}
\begin{tikzpicture}[>=triangle 45, font=\scriptsize]
\node[fill,circle,inner sep=0pt,minimum size=5pt,label={left:{$A_1$}}] (A1) at (0,0) {};
\node[fill,circle,inner sep=0pt,minimum size=5pt,label={left:{$Y_0$}}] (Y0) at (-2,0) {};
\node[fill,circle,inner sep=0pt,minimum size=5pt,label={right:{$Y_1$}}] (Y1) at (2,0) {};
\node[draw,circle,inner sep=0pt,minimum size=5pt,label={left:{$V_0$}}] (V0) at (-2.25,1) {};
\node[draw,circle,inner sep=0pt,minimum size=5pt,label={left:{$S_0$}}] (S0) at (-2.2,1.7) {};
\node[fill,circle,inner sep=0pt,minimum size=5pt,label={right:{$\Delta Y$}}] (DELTAY) at (2,-1) {};
\node[fill,circle,inner sep=0pt,minimum size=5pt,label={[xshift=0.24cm,yshift=-0.57cm]:{$W_0$}}] (W0) at (-2.3,-1.5) {};
\node[minimum size=5pt,label={left:{$e$}}] (e) at (0.25,1.4) {};
\node[minimum size=5pt,label={left:{$c$}}] (c) at (0.25,1) {};
\node[minimum size=5pt,label={left:{$c$}}] (c) at (-2,0.6) {};
\node[minimum size=5pt,label={left:{$a$}}] (a) at (1.15,0.15) {};
\node[minimum size=5pt,label={left:{$d$}}] (d) at (-2,1.35) {};
\node[minimum size=5pt,label={left:{$b$}}] (b) at (-0.7,0.55) {};
\node[minimum size=5pt,label={left:{$1$}}] (1) at (2.5,-0.4) {};
\node[minimum size=5pt,label={left:{$-1$}}] (-1) at (-0.5,-0.4) {};
\node[minimum size=5pt,label={left:{$f$}}] (f) at (0.1,-1.45) {};
\node[minimum size=5pt,label={left:{$g$}}] (g) at (-0.45,-1.05) {};
\node[minimum size=5pt,label={left:{$f$}}] (f) at (-2,-0.85) {};
\node[minimum size=5pt,label={left:{$i$}}] (i) at (-2,-2.11) {};
\node[minimum size=5pt,label={left:{$1$}}] (1) at (-0.8,-2.7) {};
\node[minimum size=5pt,label={left:{$h$}}] (h) at (1.65,-1.7) {};
\node[minimum size=5pt,label={left:{$h$}}] (h) at (-2.55,-1.2) {};
\node[fill,circle,inner sep=0pt,minimum size=5pt,label={below:{$Z_0$}}] (Z0) at (-2.6,-2.5) {};
\node[minimum size=5pt,label={left:{$j$}}] (j) at (-0.8,-3.5) {};
\node[fill,circle,inner sep=0pt,minimum size=5pt,label={below:{$Z_1$}}] (Z1) at (0.5,-2.5) {};
\node[fill,circle,inner sep=0pt,minimum size=5pt,label={below:{$Q_0$}}] (Q0) at (-2.6,-3.5) {};
\draw[->,shorten >= 1pt] (A1)--(Y1);
\draw[->,shorten >= 1pt] (V0)--(Y0);
\draw[-{Latex[length=6.5pt]}] (Y0) to[bend right=19] (DELTAY);
\draw[->,shorten >= 1pt] (Y1)--(DELTAY);
\draw[-{Latex[length=6.5pt]}] (V0) to[bend left=20] (A1);
\draw[-{Latex[length=6.5pt]}] (V0) to[bend left=15] (Y1);
\draw[->,shorten >= 1pt] (S0)--(Y0);
\draw[-{Latex[length=6.5pt]}] (S0) to[bend left=15] (Y1);
\draw[->,shorten >= 1pt] (W0)--(Y0);
\draw[-{Latex[length=6.5pt]}] (W0) to[bend right=17] (Y1);
\draw[-{Latex[length=6.5pt]}] (W0) to[bend right=20] (A1);
\draw[->,shorten >= 1pt] (Z0)--(W0);
\draw[->,shorten >= 1pt] (Z0)--(Z1);
\draw[-{Latex[length=6.5pt]}] (Z1) to[bend right=10] (Y1);
\draw[-{Latex[length=6.5pt]}] (Z0) to[bend left=30] (Y0);
\draw[-{Latex[length=6.5pt]}] (Q0) to[bend right=15] (Z1);
\end{tikzpicture}
\caption{ }
\end{subfigure}
\hspace{2cm}
\begin{subfigure}{.4\textwidth}
\centering
\begin{tikzpicture}[>=triangle 45, font=\scriptsize]
\node[fill,circle,inner sep=0pt,minimum size=5pt,label={left:{$A_1$}}] (A1) at (0,0) {};
\node[fill,circle,inner sep=0pt,minimum size=5pt,label={right:{$\Delta Y$}}] (DELTAY) at (2,0) {};
\node[draw,circle,inner sep=0pt,minimum size=5pt,label={left:{$V_0$}}] (V0) at (-0.2,1) {};
\node[draw,circle,inner sep=0pt,minimum size=5pt,label={left:{$S_0$}}] (S0) at (1.8,1) {};
\node[fill,circle,inner sep=0pt,minimum size=5pt,label={left:{$W_0$}}] (W0) at (-0.25,-1) {};
\node[fill,circle,inner sep=0pt,minimum size=5pt,label={below:{$Z_0$}}] (Z0) at (-0.55,-2) {};
\node[fill,circle,inner sep=0pt,minimum size=5pt,label={below:{$Z_1$}}] (Z1) at (1.5,-2) {};
\node[minimum size=5pt,label={left:{$a$}}] (a) at (1.25,0.15) {};
\node[minimum size=5pt,label={left:{$e-d$}}] (e-d) at (2.85,0.6) {};
\node[minimum size=5pt,label={left:{$b$}}] (b) at (0,0.6) {};
\node[minimum size=5pt,label={left:{$g$}}] (g) at (0,-0.6) {};
\node[minimum size=5pt,label={left:{$i$}}] (i) at (-0.3,-1.55) {};
\node[minimum size=5pt,label={left:{$h$}}] (h) at (2.45,-1.15) {};
\node[minimum size=5pt,label={left:{$1$}}] (1) at (0.7,-2.2) {};
\node[minimum size=5pt,label={left, rotate=24:{$-h$}}] (-h) at (1.5,-1.3) {};
\node[fill,circle,inner sep=0pt,minimum size=5pt,label={below:{$Q_0$}}] (Q0) at (-0.55,-3) {};
\node[minimum size=5pt,label={left:{$j$}}] (j) at (0.9,-2.9) {};
\draw[->,shorten >= 1pt] (A1)--(DELTAY);
\draw[->,shorten >= 1pt] (V0)--(A1);
\draw[->,shorten >= 1pt] (S0)--(DELTAY);
\draw[->,shorten >= 1pt] (W0)--(A1);
\draw[->,shorten >= 1pt] (Z0)--(W0);
\draw[-{Latex[length=6.5pt]}] (Z0) to[bend right=20] (DELTAY);
\draw[-{Latex[length=6.5pt]}] (Z1) to[bend right=20] (DELTAY);
\draw[->,shorten >= 1pt] (Z0)--(Z1);
\draw[-{Latex[length=6.5pt]}] (Q0) to[bend right=15] (Z1);
\end{tikzpicture}
\vspace{0.5cm}
\caption{ }
\end{subfigure}
\caption{Scenario with a time-invariant covariate, $W_0$, and a time-varying covariate, $Z_t$, that evolves due to a shock, $Q_0$. \textbf{(a)} Natural representation. \textbf{(b)} Compact representation. Illustrative linear model coefficients are shown as lowercase letters on each edge.}
\label{cdscenario1}
\end{figure}
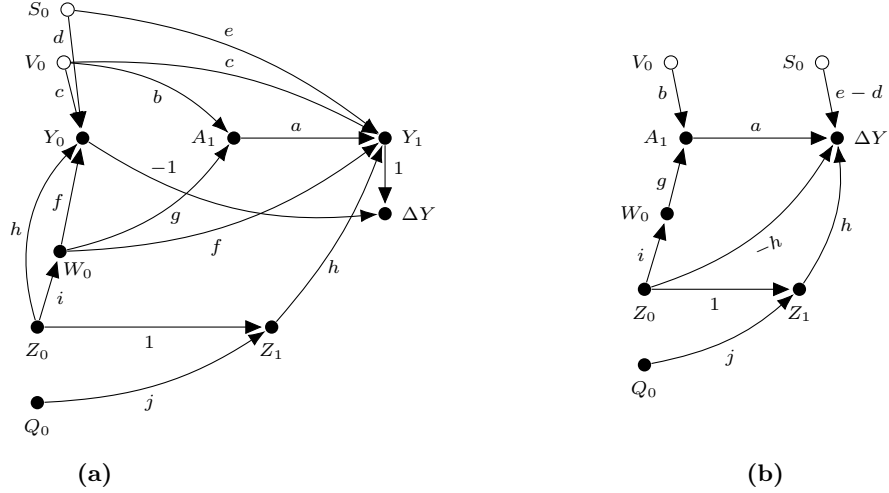

Figure \ref{cdscenario1} illustrates a setting with two time-invariant covariates, $W_0$ and $Q_0$, and a time-varying covariate, $Z_t$. In Panel (a), the evolution of $Z_t$ is driven by $Q_0$; absent this exogenous shock, $Z_t$ would remain constant, as reflected by the unit coefficient from $Z_0$ to $Z_1$. Both $W_0$ and $Z_t$ have time-invariant direct effects, $f$ and $h$, respectively, on the outcome over time. Panel (b) shows the corresponding compact representation. Given that $\sigma(\Delta Y, W_0)=0$, the edge from $W_0$ to $\Delta Y$ is removed. In contrast, $\sigma(\Delta Y, Z_0)=iga \neq 0$. Since the indirect paths $Z_0 \rightarrow W_0 \rightarrow A_1 \rightarrow \Delta Y$ with magnitude $iga$ and $Z_0 \rightarrow Z_1 \rightarrow \Delta Y$ with magnitude $h$ are retained in the compact representation, a direct edge from $Z_0$ to $\Delta Y$ with magnitude $-h$ is included to preserve the covariance structure. 

Although there are backdoor paths from $A_1$ to $\Delta Y$ in the compact representation, the time symmetry of the confounding ensures perfect offsetting in \ac{DiD} such that $\sigma(\Delta Y,A_1)=a+gih-gih=a$. Thus, unconditional parallel trends (or equi-confounding) holds. This example is intentionally simple, but illustrates that parallel trends requires \emph{all} confounding to cancel out exactly. As the number of covariates increases, we may find a causal model with such perfect offsetting implausible. A more plausible model may include backdoor paths that break the time symmetry of confounding, in which case conditioning on covariates become necessary.

That said, \ac{DiD} always assumes perfect symmetry of \emph{unmeasured} confounding, an assumption that may also be hard to justify, though it can still be preferable when the anticipated bias is smaller than that under alternative identification strategies. The important takeaway is that it is not necessary to assume that all confounding is perfectly offset. Adjusting for covariates under conditional parallel trends relaxes the reliance on perfect offsetting of \emph{all} confounding. However, conditioning on covariates can also break the symmetry and paradoxically introduce bias. In Figure \ref{cdscenario1}, adjusting for $Z_1$ alone would break parallel trends because the backdoor path through $Z_0$ would remain open and the confounding along that path would no longer be offset by the confounding through $Z_1$. While adjusting for $W_0$ and both values of $Z_t$ under conditional parallel trends yields the same result as an unconditional analysis in this context, indiscriminate covariate adjustment can introduce problems such as collider bias or limited positivity in other contexts.

In practice, it is difficult to justify assumptions about the stability of all covariate effects over time, and thus we generally cannot be certain that all confounding operating through observed covariates is perfectly offset. Therefore, the choice between unconditional or conditional parallel trends should be guided by the relationships encoded in the causal model. As noted above, the structural coefficients retained in the causal diagrams that follow are included only to clarify the mechanics of each scenario, not as a linearity requirement.

\subsection{Time-invariant covariates for time-varying confounding}\label{ss:time-varying-effect}

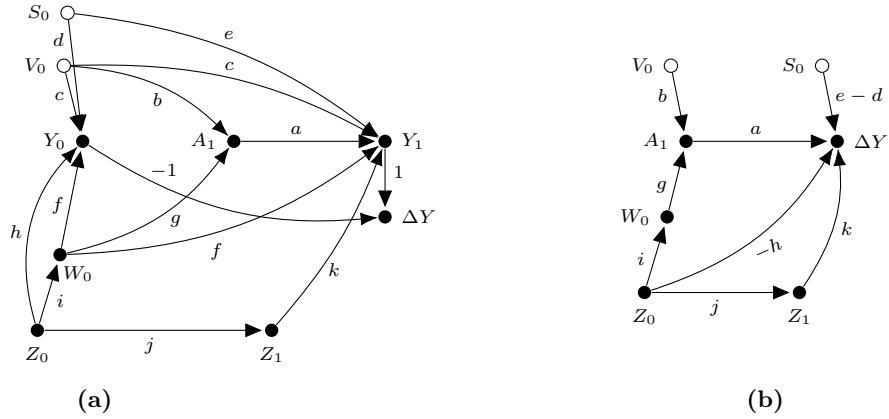
\begin{figure}[h!]
\begin{subfigure}{.4\textwidth}
\hspace{2cm}
\begin{tikzpicture}[>=triangle 45, font=\scriptsize]
\node[fill,circle,inner sep=0pt,minimum size=5pt,label={left:{$A_1$}}] (A1) at (0,0) {};
\node[fill,circle,inner sep=0pt,minimum size=5pt,label={left:{$Y_0$}}] (Y0) at (-2,0) {};
\node[fill,circle,inner sep=0pt,minimum size=5pt,label={right:{$Y_1$}}] (Y1) at (2,0) {};
\node[draw,circle,inner sep=0pt,minimum size=5pt,label={left:{$V_0$}}] (V0) at (-2.25,1) {};
\node[draw,circle,inner sep=0pt,minimum size=5pt,label={left:{$S_0$}}] (S0) at (-2.2,1.7) {};
\node[fill,circle,inner sep=0pt,minimum size=5pt,label={right:{$\Delta Y$}}] (DELTAY) at (2,-1) {};
\node[fill,circle,inner sep=0pt,minimum size=5pt,label={[xshift=0.24cm,yshift=-0.57cm]:{$W_0$}}] (W0) at (-2.3,-1.5) {};
\node[minimum size=5pt,label={left:{$e$}}] (e) at (0.25,1.4) {};
\node[minimum size=5pt,label={left:{$c$}}] (c) at (0.25,1) {};
\node[minimum size=5pt,label={left:{$c$}}] (c) at (-2,0.6) {};
\node[minimum size=5pt,label={left:{$a$}}] (a) at (1.15,0.15) {};
\node[minimum size=5pt,label={left:{$d$}}] (d) at (-2,1.35) {};
\node[minimum size=5pt,label={left:{$b$}}] (b) at (-0.7,0.55) {};
\node[minimum size=5pt,label={left:{$1$}}] (1) at (2.5,-0.4) {};
\node[minimum size=5pt,label={left:{$-1$}}] (-1) at (-0.5,-0.4) {};
\node[minimum size=5pt,label={left:{$f$}}] (f) at (0.1,-1.45) {};
\node[minimum size=5pt,label={left:{$g$}}] (g) at (-0.45,-1.05) {};
\node[minimum size=5pt,label={left:{$f$}}] (f) at (-2,-0.85) {};
\node[minimum size=5pt,label={left:{$i$}}] (i) at (-2,-2.11) {};
\node[minimum size=5pt,label={left:{$j$}}] (j) at (-0.8,-2.7) {};
\node[minimum size=5pt,label={left:{$k$}}] (k) at (1.65,-1.7) {};
\node[minimum size=5pt,label={left:{$h$}}] (h) at (-2.55,-1.2) {};
\node[fill,circle,inner sep=0pt,minimum size=5pt,label={below:{$Z_0$}}] (Z0) at (-2.6,-2.5) {};
\node[fill,circle,inner sep=0pt,minimum size=5pt,label={below:{$Z_1$}}] (Z1) at (0.5,-2.5) {};
\draw[->,shorten >= 1pt] (A1)--(Y1);
\draw[->,shorten >= 1pt] (V0)--(Y0);
\draw[-{Latex[length=6.5pt]}] (Y0) to[bend right=19] (DELTAY);
\draw[->,shorten >= 1pt] (Y1)--(DELTAY);
\draw[-{Latex[length=6.5pt]}] (V0) to[bend left=20] (A1);
\draw[-{Latex[length=6.5pt]}] (V0) to[bend left=15] (Y1);
\draw[->,shorten >= 1pt] (S0)--(Y0);
\draw[-{Latex[length=6.5pt]}] (S0) to[bend left=15] (Y1);
\draw[->,shorten >= 1pt] (W0)--(Y0);
\draw[-{Latex[length=6.5pt]}] (W0) to[bend right=17] (Y1);
\draw[-{Latex[length=6.5pt]}] (W0) to[bend right=20] (A1);
\draw[->,shorten >= 1pt] (Z0)--(W0);
\draw[->,shorten >= 1pt] (Z0)--(Z1);
\draw[-{Latex[length=6.5pt]}] (Z1) to[bend right=10] (Y1);
\draw[-{Latex[length=6.5pt]}] (Z0) to[bend left=30] (Y0);
\end{tikzpicture}
\caption{ }
\end{subfigure}
\hspace{2cm}
\begin{subfigure}{.4\textwidth}
\centering
\begin{tikzpicture}[>=triangle 45, font=\scriptsize]
\node[fill,circle,inner sep=0pt,minimum size=5pt,label={left:{$A_1$}}] (A1) at (0,0) {};
\node[fill,circle,inner sep=0pt,minimum size=5pt,label={right:{$\Delta Y$}}] (DELTAY) at (2,0) {};
\node[draw,circle,inner sep=0pt,minimum size=5pt,label={left:{$V_0$}}] (V0) at (-0.2,1) {};
\node[draw,circle,inner sep=0pt,minimum size=5pt,label={left:{$S_0$}}] (S0) at (1.8,1) {};
\node[fill,circle,inner sep=0pt,minimum size=5pt,label={left:{$W_0$}}] (W0) at (-0.25,-1) {};
\node[fill,circle,inner sep=0pt,minimum size=5pt,label={below:{$Z_0$}}] (Z0) at (-0.55,-2) {};
\node[fill,circle,inner sep=0pt,minimum size=5pt,label={below:{$Z_1$}}] (Z1) at (1.5,-2) {};
\node[minimum size=5pt,label={left:{$a$}}] (a) at (1.25,0.15) {};
\node[minimum size=5pt,label={left:{$e-d$}}] (e-d) at (2.85,0.6) {};
\node[minimum size=5pt,label={left:{$b$}}] (b) at (0,0.6) {};
\node[minimum size=5pt,label={left:{$g$}}] (g) at (0,-0.6) {};
\node[minimum size=5pt,label={left:{$i$}}] (i) at (-0.3,-1.55) {};
\node[minimum size=5pt,label={left:{$k$}}] (k) at (2.45,-1.15) {};
\node[minimum size=5pt,label={left:{$j$}}] (j) at (0.7,-2.2) {};
\node[minimum size=5pt,label={left, rotate=24:{$-h$}}] (-h) at (1.5,-1.3) {};
\draw[->,shorten >= 1pt] (A1)--(DELTAY);
\draw[->,shorten >= 1pt] (V0)--(A1);
\draw[->,shorten >= 1pt] (S0)--(DELTAY);
\draw[->,shorten >= 1pt] (W0)--(A1);
\draw[->,shorten >= 1pt] (Z0)--(W0);
\draw[-{Latex[length=6.5pt]}] (Z0) to[bend right=20] (DELTAY);
\draw[-{Latex[length=6.5pt]}] (Z1) to[bend right=20] (DELTAY);
\draw[->,shorten >= 1pt] (Z0)--(Z1);
\end{tikzpicture}
\vspace{0.5cm}
\caption{ }
\end{subfigure}
\caption{Scenario with a time-invariant covariate $W_0$ and a time-varying covariate $Z_t$. \textbf{(a)} Natural representation. \textbf{(b)} Compact representation. Illustrative linear model coefficients are shown as lowercase letters on each edge.}
\label{cdscenario2}
\end{figure}

Next, we show how even a variable that would not be considered a ``\ac{DiD} confounder'' can be useful as a conditioning variable. Figure \ref{cdscenario2} depicts a setting with one time-invariant covariate, $W_0$, with a time-invariant effect $f$ on the outcome, and a time-varying covariate, $Z_t$ with a time-varying effect ($h \neq k$). Panel (a) shows the natural representation and Panel (b) the compact representation, in which the edge from $W_0$ to $\Delta Y$ is again absent since its effects cancel, while the edges from $Z_0$ and $Z_1$ to $\Delta Y$ are retained.

In this case, unconditional parallel trends no longer holds, $\sigma(\Delta Y, A_1) = a+gijk-gih \neq a$, which means we should resort to conditional parallel trends. By applying the backdoor criterion to the compact graph in Panel (b) implies that either $\{Z_0\}$ or $\{W_0\}$ is minimally sufficient, since neither is a descendant of treatment, and each is sufficient to block all backdoor paths from $A_1$ to $\Delta Y$.

Although $W_0$ is time-invariant and has a time-invariant effect on the outcome, and thus would not traditionally be considered a \ac{DiD} confounder, conditioning on it nonetheless enables \ac{DiD} identification via conditional parallel trends. This example highlights that selecting adjustment sets by reasoning separately about whether variables change over time or have time-varying effects on the outcome may overlook useful conditioning variables. In particular, a variable like $W_0$ can be leveraged when $Z_0$ is unmeasured or measured with error.

\subsection{Adjustment for post-treatment covariates}\label{ss:post-trt-vars}
 
Conventional guidance cautions against adjusting for post-treatment covariates in \ac{DiD} studies because of concern that these variables might be mediators or colliders \cite{stuartUsingPropensityScores2014a}.
In this section, we present four scenarios that share a common feature: a time-varying covariate with a time-varying effect on the outcome, so that unconditional parallel trends does not hold in any of them. The scenarios differ in the mechanism driving the covariate's evolution: an auto-regressive process (Figure \ref{cdscenario3a}), an endogenous shock (Figure \ref{cdscenario3b}), an unmeasured confounder (Figure \ref{cdscenario3c}) or the treatment itself (Figure \ref{cdscenario3d}). Through these examples, we illustrate how features of the causal model guide whether post-treatment covariates should be conditioned on.

\subsubsection{Auto-regressive process}

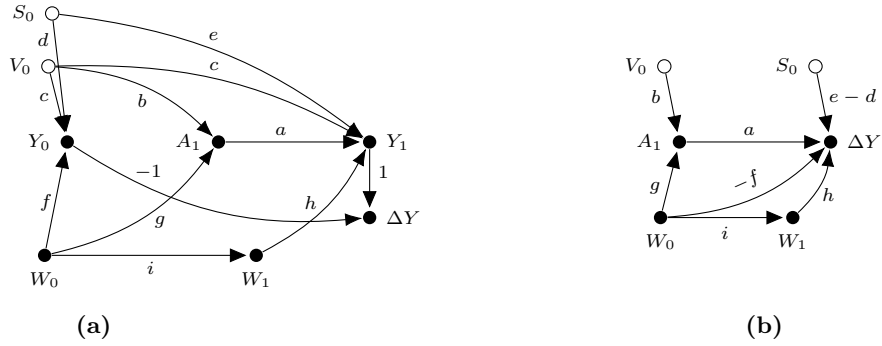
\begin{figure}[h!]
\begin{subfigure}{.4\textwidth}
\hspace{2cm}
\begin{tikzpicture}[>=triangle 45, font=\scriptsize]
\node[fill,circle,inner sep=0pt,minimum size=5pt,label={left:{$A_1$}}] (A1) at (0,0) {};
\node[fill,circle,inner sep=0pt,minimum size=5pt,label={left:{$Y_0$}}] (Y0) at (-2,0) {};
\node[fill,circle,inner sep=0pt,minimum size=5pt,label={right:{$Y_1$}}] (Y1) at (2,0) {};
\node[draw,circle,inner sep=0pt,minimum size=5pt,label={left:{$V_0$}}] (V0) at (-2.25,1) {};
\node[draw,circle,inner sep=0pt,minimum size=5pt,label={left:{$S_0$}}] (S0) at (-2.2,1.7) {};
\node[fill,circle,inner sep=0pt,minimum size=5pt,label={right:{$\Delta Y$}}] (DELTAY) at (2,-1) {};
\node[fill,circle,inner sep=0pt,minimum size=5pt,label={below:{$W_0$}}] (W0) at (-2.3,-1.5) {};
\node[fill,circle,inner sep=0pt,minimum size=5pt,label={below:{$W_1$}}] (W1) at (0.5,-1.5) {};
\node[minimum size=5pt,label={left:{$e$}}] (e) at (0.25,1.4) {};
\node[minimum size=5pt,label={left:{$c$}}] (c) at (0.25,1) {};
\node[minimum size=5pt,label={left:{$c$}}] (c) at (-2,0.6) {};
\node[minimum size=5pt,label={left:{$a$}}] (a) at (1.15,0.15) {};
\node[minimum size=5pt,label={left:{$d$}}] (d) at (-2,1.35) {};
\node[minimum size=5pt,label={left:{$b$}}] (b) at (-0.7,0.55) {};
\node[minimum size=5pt,label={left:{$1$}}] (1) at (2.5,-0.4) {};
\node[minimum size=5pt,label={left:{$-1$}}] (-1) at (-0.5,-0.4) {};
\node[minimum size=5pt,label={left:{$h$}}] (h) at (1.55,-0.8) {};
\node[minimum size=5pt,label={left:{$g$}}] (g) at (-0.45,-1.05) {};
\node[minimum size=5pt,label={left:{$f$}}] (f) at (-1.95,-0.8) {};
\node[minimum size=5pt,label={left:{$i$}}] (i) at (-0.6,-1.67) {};
\draw[->,shorten >= 1pt] (A1)--(Y1);
\draw[->,shorten >= 1pt] (V0)--(Y0);
\draw[-{Latex[length=6.5pt]}] (Y0) to[bend right=19] (DELTAY);
\draw[->,shorten >= 1pt] (Y1)--(DELTAY);
\draw[-{Latex[length=6.5pt]}] (V0) to[bend left=20] (A1);
\draw[-{Latex[length=6.5pt]}] (V0) to[bend left=15] (Y1);
\draw[->,shorten >= 1pt] (S0)--(Y0);
\draw[-{Latex[length=6.5pt]}] (S0) to[bend left=15] (Y1);
\draw[->,shorten >= 1pt] (W0)--(Y0);
\draw[-{Latex[length=6.5pt]}] (W0) to[bend right=20] (A1);
\draw[->,shorten >= 1pt] (W0)--(W1);
\draw[-{Latex[length=6.5pt]}] (W1) to[bend right=17] (Y1);
\end{tikzpicture}
\caption{ }
\end{subfigure}
\hspace{2cm}
\begin{subfigure}{.4\textwidth}
\centering
\begin{tikzpicture}[>=triangle 45, font=\scriptsize]
\node[fill,circle,inner sep=0pt,minimum size=5pt,label={left:{$A_1$}}] (A1) at (0,0) {};
\node[fill,circle,inner sep=0pt,minimum size=5pt,label={right:{$\Delta Y$}}] (DELTAY) at (2,0) {};
\node[draw,circle,inner sep=0pt,minimum size=5pt,label={left:{$V_0$}}] (V0) at (-0.2,1) {};
\node[draw,circle,inner sep=0pt,minimum size=5pt,label={left:{$S_0$}}] (S0) at (1.8,1) {};
\node[fill,circle,inner sep=0pt,minimum size=5pt,label={below:{$W_0$}}] (W0) at (-0.25,-1) {};
\node[fill,circle,inner sep=0pt,minimum size=5pt,label={below:{$W_1$}}] (W1) at (1.5,-1) {};
\node[minimum size=5pt,label={left:{$a$}}] (a) at (1.25,0.15) {};
\node[minimum size=5pt,label={left:{$e-d$}}] (e-d) at (2.85,0.6) {};
\node[minimum size=5pt,label={left:{$b$}}] (b) at (0,0.6) {};
\node[minimum size=5pt,label={left:{$g$}}] (g) at (0,-0.6) {};
\node[minimum size=5pt,label={left, rotate=24:{$-f$}}] (-f) at (1.3,-0.4) {};
\node[minimum size=5pt,label={left:{$h$}}] (h) at (2.3,-0.7) {};
\node[minimum size=5pt,label={left:{$i$}}] (i) at (0.9,-1.2) {};
\draw[->,shorten >= 1pt] (A1)--(DELTAY);
\draw[->,shorten >= 1pt] (V0)--(A1);
\draw[->,shorten >= 1pt] (S0)--(DELTAY);
\draw[->,shorten >= 1pt] (W0)--(A1);
\draw[->,shorten >= 1pt] (W0)--(W1);
\draw[-{Latex[length=6.5pt]}] (W0) to[bend right=20] (DELTAY);
\draw[-{Latex[length=6.5pt]}] (W1) to[bend right=20] (DELTAY);
\end{tikzpicture}
\vspace{0.5cm}
\caption{ }
\end{subfigure}
\caption{Scenario in which the time-varying covariate $W_t$'s evolution is governed by an auto-regressive process. \textbf{(a)} Natural diagram. \textbf{(b)} Compact representation. Illustrative linear model coefficients are shown as lowercase letters on each edge.}
\label{cdscenario3a}
\end{figure}

In Figure \ref{cdscenario3a}, the only cause of the covariate's evolution is the previous value of the covariate, $W_0 \rightarrow W_1$. From the compact representation in Panel (b), we can see that $\{W_0\}$ is minimally sufficient to block backdoor paths from $A_1$ to $\Delta Y$. We could also adjust for both $W_0$ and $W_1$ without adverse consequences, since $W_1$ is not a descendant of treatment. However, $W_1$ alone is not sufficient as it cannot block all backdoor paths.

This scenario illustrates a case in which adjustment for the post-treatment covariate is not required for identification, but can be included without adverse consequences.

\subsubsection{Endogenous shock}

\begin{figure}[h!]
\begin{subfigure}{.4\textwidth}
\hspace{2cm}
\begin{tikzpicture}[>=triangle 45, font=\scriptsize]
\node[fill,circle,inner sep=0pt,minimum size=5pt,label={left:{$A_1$}}] (A1) at (0,0) {};
\node[fill,circle,inner sep=0pt,minimum size=5pt,label={left:{$Y_0$}}] (Y0) at (-2,0) {};
\node[fill,circle,inner sep=0pt,minimum size=5pt,label={right:{$Y_1$}}] (Y1) at (2,0) {};
\node[draw,circle,inner sep=0pt,minimum size=5pt,label={left:{$V_0$}}] (V0) at (-2.25,1) {};
\node[draw,circle,inner sep=0pt,minimum size=5pt,label={left:{$S_0$}}] (S0) at (-2.2,1.7) {};
\node[fill,circle,inner sep=0pt,minimum size=5pt,label={right:{$\Delta Y$}}] (DELTAY) at (2,-1) {};
\node[fill,circle,inner sep=0pt,minimum size=5pt,label={below:{$W_0$}}] (W0) at (-2.3,-1.5) {};
\node[fill,circle,inner sep=0pt,minimum size=5pt,label={[xshift=-9pt,yshift=-9pt]1:{$W_1$}}] (W1) at (0.5,-1.5) {};
\node[fill,circle,inner sep=0pt,minimum size=5pt,label={below:{$Z_0$}}] (Z0) at (-2.3,-2.5) {};
\node[minimum size=5pt,label={left:{$e$}}] (e) at (0.25,1.4) {};
\node[minimum size=5pt,label={left:{$c$}}] (c) at (0.25,1) {};
\node[minimum size=5pt,label={left:{$c$}}] (c) at (-2,0.6) {};
\node[minimum size=5pt,label={left:{$a$}}] (a) at (1.15,0.15) {};
\node[minimum size=5pt,label={left:{$d$}}] (d) at (-2,1.35) {};
\node[minimum size=5pt,label={left:{$b$}}] (b) at (-0.7,0.55) {};
\node[minimum size=5pt,label={left:{$1$}}] (1) at (2.5,-0.4) {};
\node[minimum size=5pt,label={left:{$-1$}}] (-1) at (-0.5,-0.4) {};
\node[minimum size=5pt,label={left:{$h$}}] (h) at (1.55,-0.75) {};
\node[minimum size=5pt,label={left:{$g$}}] (g) at (-0.45,-1.05) {};
\node[minimum size=5pt,label={left:{$f$}}] (f) at (-1.95,-0.8) {};
\node[minimum size=5pt,label={left:{$i$}}] (i) at (0.05,-1.3) {};
\node[minimum size=5pt,label={left:{$k$}}] (k) at (-0.7,-1.67) {};
\node[minimum size=5pt,label={left:{$j$}}] (j) at (-0.65,-2.5) {};
\draw[->,shorten >= 1pt] (A1)--(Y1);
\draw[->,shorten >= 1pt] (V0)--(Y0);
\draw[-{Latex[length=6.5pt]}] (Y0) to[bend right=19] (DELTAY);
\draw[->,shorten >= 1pt] (Y1)--(DELTAY);
\draw[-{Latex[length=6.5pt]}] (V0) to[bend left=20] (A1);
\draw[-{Latex[length=6.5pt]}] (V0) to[bend left=15] (Y1);
\draw[->,shorten >= 1pt] (S0)--(Y0);
\draw[-{Latex[length=6.5pt]}] (S0) to[bend left=15] (Y1);
\draw[->,shorten >= 1pt] (W0)--(Y0);
\draw[-{Latex[length=6.5pt]}] (W0) to[bend right=20] (A1);
\draw[->,shorten >= 1pt] (W0)--(W1);
\draw[-{Latex[length=6.5pt]}] (W1) to[bend right=13] (Y1);
\draw[-{Latex[length=6.5pt]}] (Z0) to[bend right=17] (W1);
\draw[-{Latex[length=6.5pt]}] (Z0) to[bend right=35] (A1);
\end{tikzpicture}
\caption{ }
\end{subfigure}
\hspace{2cm}
\begin{subfigure}{.4\textwidth}
\centering
\begin{tikzpicture}[>=triangle 45, font=\scriptsize]
\node[fill,circle,inner sep=0pt,minimum size=5pt,label={left:{$A_1$}}] (A1) at (0,0) {};
\node[fill,circle,inner sep=0pt,minimum size=5pt,label={right:{$\Delta Y$}}] (DELTAY) at (2,0) {};
\node[draw,circle,inner sep=0pt,minimum size=5pt,label={left:{$V_0$}}] (V0) at (-0.2,1) {};
\node[draw,circle,inner sep=0pt,minimum size=5pt,label={left:{$S_0$}}] (S0) at (1.8,1) {};
\node[fill,circle,inner sep=0pt,minimum size=5pt,label={left:{$W_0$}}] (W0) at (-0.25,-1) {};
\node[fill,circle,inner sep=0pt,minimum size=5pt,label={below:{$W_1$}}] (W1) at (1.5,-1) {};
\node[minimum size=5pt,label={left:{$a$}}] (a) at (1.25,0.15) {};
\node[minimum size=5pt,label={left:{$e-d$}}] (e-d) at (2.85,0.6) {};
\node[minimum size=5pt,label={left:{$b$}}] (b) at (0,0.6) {};
\node[minimum size=5pt,label={left:{$g$}}] (g) at (0,-0.6) {};
\node[minimum size=5pt,label={left, rotate=24:{$-f$}}] (-f) at (1.3,-0.4) {};
\node[minimum size=5pt,label={left:{$i$}}] (i) at (0.5,-1.4) {};
\node[minimum size=5pt,label={left:{$h$}}] (h) at (2.3,-0.7) {};
\node[minimum size=5pt,label={left:{$k$}}] (k) at (0.9,-1.17) {};
\node[minimum size=5pt,label={left:{$j$}}] (j) at (1,-1.8) {};
\node[fill,circle,inner sep=0pt,minimum size=5pt,label={below:{$Z_0$}}] (Z0) at (-0.25,-2) {};
\draw[->,shorten >= 1pt] (A1)--(DELTAY);
\draw[->,shorten >= 1pt] (V0)--(A1);
\draw[->,shorten >= 1pt] (S0)--(DELTAY);
\draw[->,shorten >= 1pt] (W0)--(A1);
\draw[->,shorten >= 1pt] (W0)--(W1);
\draw[-{Latex[length=6.5pt]}] (W0) to[bend right=20] (DELTAY);
\draw[-{Latex[length=6.5pt]}] (W1) to[bend right=20] (DELTAY);
\draw[-{Latex[length=6.5pt]}] (Z0) to[bend right=10] (W1);
\draw[-{Latex[length=6.5pt]}] (Z0) to[bend right=25] (A1);
\end{tikzpicture}
\vspace{0.5cm}
\caption{ }
\end{subfigure}
\caption{Scenario in which the time-varying covariate $W_t$'s evolution is governed by observed confounder $Z_0$ and an auto-regressive process. \textbf{(a)} Natural diagram. \textbf{(b)} Compact representation. Illustrative linear model coefficients are shown as lowercase letters on each edge.}
\label{cdscenario3b}
\end{figure}
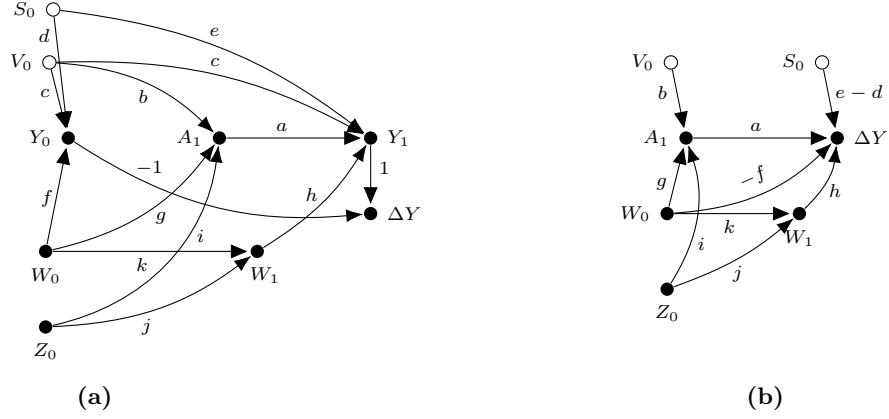

In Figure \ref{cdscenario3b}, the evolution of $W_t$ is driven by both an endogenous shock, $Z_0$, and an auto-regressive process. Applying the backdoor criterion to the compact representation in Panel (b), we find that either $\{W_0,Z_0\}$ or $\{W_0,W_1\}$ are minimally sufficient to block all backdoor paths from $A_1$ to $\Delta Y$. This is the first scenario in which adjusting for both values of a time-varying covariate might be required for identification.

This scenario illustrates a case in which adjustment for the post-treatment covariate, alongside the pre-treatment value, is one route to identification, but not the only one.

\subsubsection{Unmeasured confounder}\label{ss:post-trt-vars3}

\begin{figure}[h!]
\begin{subfigure}{.4\textwidth}
\hspace{2cm}
\begin{tikzpicture}[>=triangle 45, font=\scriptsize]
\node[fill,circle,inner sep=0pt,minimum size=5pt,label={left:{$A_1$}}] (A1) at (0,0) {};
\node[fill,circle,inner sep=0pt,minimum size=5pt,label={left:{$Y_0$}}] (Y0) at (-2,0) {};
\node[fill,circle,inner sep=0pt,minimum size=5pt,label={right:{$Y_1$}}] (Y1) at (2,0) {};
\node[draw,circle,inner sep=0pt,minimum size=5pt,label={left:{$V_0$}}] (V0) at (-2.25,1) {};
\node[draw,circle,inner sep=0pt,minimum size=5pt,label={left:{$S_0$}}] (S0) at (-2.2,1.7) {};
\node[fill,circle,inner sep=0pt,minimum size=5pt,label={right:{$\Delta Y$}}] (DELTAY) at (2,-1) {};
\node[fill,circle,inner sep=0pt,minimum size=5pt,label={below:{$W_0$}}] (W0) at (-2.3,-1.5) {};
\node[fill,circle,inner sep=0pt,minimum size=5pt,label={below:{$W_1$}}] (W1) at (0.5,-1.5) {};
\node[minimum size=5pt,label={left:{$e$}}] (e) at (0.25,1.4) {};
\node[minimum size=5pt,label={left:{$c$}}] (c) at (0.25,1) {};
\node[minimum size=5pt,label={left:{$c$}}] (c) at (-2,0.6) {};
\node[minimum size=5pt,label={left:{$a$}}] (a) at (1.15,0.15) {};
\node[minimum size=5pt,label={left:{$d$}}] (d) at (-2,1.35) {};
\node[minimum size=5pt,label={left:{$b$}}] (b) at (-0.7,0.55) {};
\node[minimum size=5pt,label={left:{$1$}}] (1) at (2.5,-0.4) {};
\node[minimum size=5pt,label={left:{$-1$}}] (-1) at (1.75,-1.25) {};
\node[minimum size=5pt,label={left:{$h$}}] (h) at (1.55,-0.8) {};
\node[minimum size=5pt,label={left:{$g$}}] (g) at (-0.45,-1.05) {};
\node[minimum size=5pt,label={left:{$f$}}] (f) at (-1.95,-0.8) {};
\node[minimum size=5pt,label={left:{$j$}}] (j) at (-0.65,-1.7) {};
\node[minimum size=5pt,label={left:{$i$}}] (i) at (-0.79,-0.1) {};
\draw[->,shorten >= 1pt] (A1)--(Y1);
\draw[->,shorten >= 1pt] (V0)--(Y0);
\draw[-{Latex[length=6.5pt]}] (Y0) to[bend right=19] (DELTAY);
\draw[->,shorten >= 1pt] (Y1)--(DELTAY);
\draw[-{Latex[length=6.5pt]}] (V0) to[bend left=20] (A1);
\draw[-{Latex[length=6.5pt]}] (V0) to[bend left=15] (Y1);
\draw[->,shorten >= 1pt] (S0)--(Y0);
\draw[-{Latex[length=6.5pt]}] (S0) to[bend left=15] (Y1);
\draw[->,shorten >= 1pt] (W0)--(Y0);
\draw[-{Latex[length=6.5pt]}] (W0) to[bend right=20] (A1);
\draw[->,shorten >= 1pt] (W0)--(W1);
\draw[-{Latex[length=6.5pt]}] (W1) to[bend right=17] (Y1);
\draw[-{Latex[length=6.5pt]}] (V0) to[bend right=12] (W1);
\end{tikzpicture}
\caption{ }
\end{subfigure}
\hspace{3cm}
\begin{subfigure}{.4\textwidth}
\centering
\begin{tikzpicture}[>=triangle 45, font=\scriptsize]
\node[fill,circle,inner sep=0pt,minimum size=5pt,label={left:{$A_1$}}] (A1) at (0,0) {};
\node[fill,circle,inner sep=0pt,minimum size=5pt,label={right:{$\Delta Y$}}] (DELTAY) at (2,0) {};
\node[draw,circle,inner sep=0pt,minimum size=5pt,label={left:{$V_0$}}] (V0) at (-0.2,1) {};
\node[draw,circle,inner sep=0pt,minimum size=5pt,label={left:{$S_0$}}] (S0) at (1.8,1) {};
\node[fill,circle,inner sep=0pt,minimum size=5pt,label={below:{$W_0$}}] (W0) at (-0.25,-1) {};
\node[fill,circle,inner sep=0pt,minimum size=5pt,label={below:{$W_1$}}] (W1) at (1.5,-1) {};
\node[minimum size=5pt,label={left:{$a$}}] (a) at (1.25,0.15) {};
\node[minimum size=5pt,label={left:{$e-d$}}] (e-d) at (2.85,0.6) {};
\node[minimum size=5pt,label={left:{$b$}}] (b) at (0,0.6) {};
\node[minimum size=5pt,label={left:{$g$}}] (g) at (0,-0.6) {};
\node[minimum size=5pt,label={left, rotate=24:{$-f$}}] (-f) at (0.95,-0.55) {};
\node[minimum size=5pt,label={left:{$i$}}] (i) at (0.7,0.35) {};
\node[minimum size=5pt,label={left:{$h$}}] (h) at (2.3,-0.7) {};
\node[minimum size=5pt,label={left:{$j$}}] (j) at (0.9,-1.2) {};
\draw[->,shorten >= 1pt] (A1)--(DELTAY);
\draw[->,shorten >= 1pt] (V0)--(A1);
\draw[->,shorten >= 1pt] (S0)--(DELTAY);
\draw[->,shorten >= 1pt] (W0)--(A1);
\draw[->,shorten >= 1pt] (W0)--(W1);
\draw[-{Latex[length=6.5pt]}] (W0) to[bend right=20] (DELTAY);
\draw[-{Latex[length=6.5pt]}] (W1) to[bend right=20] (DELTAY);
\draw[-{Latex[length=6.5pt]}] (V0) to[bend right=12] (W1);
\end{tikzpicture}
\vspace{0.5cm}
\caption{ }
\end{subfigure}
\caption{Scenario in which the time-varying covariate $W_t$'s evolution is governed by the unobserved confounder $V_0$ and an auto-regressive process. \textbf{(a)} Natural diagram. \textbf{(b)} Compact representation.  Illustrative linear model coefficients are shown as lowercase letters on each edge.}
\label{cdscenario3c}
\end{figure}
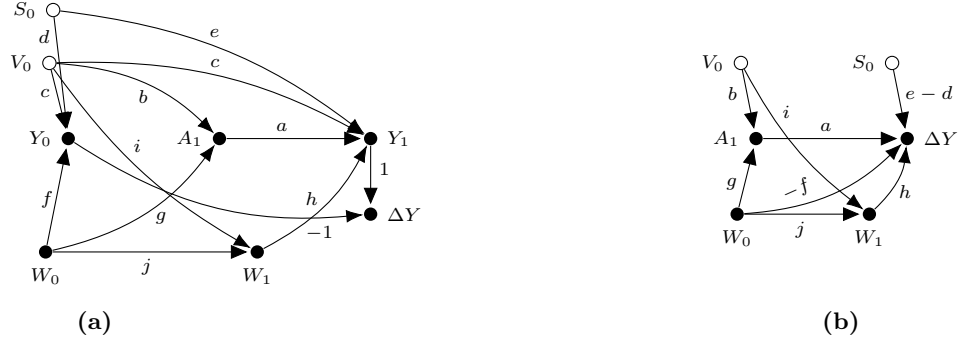

In Figure \ref{cdscenario3c}, the covariate's evolution is determined by the same unmeasured, time-invariant confounder $V_0$ that motivates the use of \ac{DiD} plus an auto-regressive dependence on $W_0$. From the compact representation in Panel (b), we can see that $\{W_0,W_1\}$ is minimally sufficient to block all backdoor paths. Given that the unmeasured confounder $V_0$ cannot be adjusted for directly, in contrast to the observed shock $Z_0$ in the previous scenario, adjustment for both values of the observed time-varying covariate $W_t$ is required for identification.

This scenario demonstrates that adjustment for the post-treatment covariate can be the only available route to identification when unobserved confounders drive the covariate's evolution.

\subsubsection{Treatment}

\begin{figure}[h!]
\begin{subfigure}{.4\textwidth}
\hspace{2cm}
\begin{tikzpicture}[>=triangle 45, font=\scriptsize]
\node[fill,circle,inner sep=0pt,minimum size=5pt,label={left:{$A_1$}}] (A1) at (0,0) {};
\node[fill,circle,inner sep=0pt,minimum size=5pt,label={left:{$Y_0$}}] (Y0) at (-2,0) {};
\node[fill,circle,inner sep=0pt,minimum size=5pt,label={right:{$Y_1$}}] (Y1) at (2,0) {};
\node[draw,circle,inner sep=0pt,minimum size=5pt,label={left:{$V_0$}}] (V0) at (-2.25,1) {};
\node[draw,circle,inner sep=0pt,minimum size=5pt,label={left:{$S_0$}}] (S0) at (-2.2,1.7) {};
\node[fill,circle,inner sep=0pt,minimum size=5pt,label={right:{$\Delta Y$}}] (DELTAY) at (2,-1) {};
\node[fill,circle,inner sep=0pt,minimum size=5pt,label={below:{$W_0$}}] (W0) at (-2.3,-1.5) {};
\node[fill,circle,inner sep=0pt,minimum size=5pt,label={below:{$W_1$}}] (W1) at (0.5,-1.5) {};
\node[minimum size=5pt,label={left:{$e$}}] (e) at (0.25,1.4) {};
\node[minimum size=5pt,label={left:{$c$}}] (c) at (0.25,1) {};
\node[minimum size=5pt,label={left:{$c$}}] (c) at (-2,0.6) {};
\node[minimum size=5pt,label={left:{$a$}}] (a) at (1.15,0.15) {};
\node[minimum size=5pt,label={left:{$d$}}] (d) at (-2,1.35) {};
\node[minimum size=5pt,label={left:{$b$}}] (b) at (-0.7,0.55) {};
\node[minimum size=5pt,label={left:{$1$}}] (1) at (2.5,-0.4) {};
\node[minimum size=5pt,label={left:{$-1$}}] (-1) at (-0.5,-0.4) {};
\node[minimum size=5pt,label={left:{$h$}}] (h) at (1.55,-0.8) {};
\node[minimum size=5pt,label={left:{$g$}}] (g) at (-0.45,-1.05) {};
\node[minimum size=5pt,label={left:{$f$}}] (f) at (-1.95,-0.8) {};
\node[minimum size=5pt,label={left:{$j$}}] (j) at (-0.65,-1.7) {};
\node[minimum size=5pt,label={left:{$i$}}] (i) at (0.7,-0.6) {};
\draw[->,shorten >= 1pt] (A1)--(Y1);
\draw[->,shorten >= 1pt] (V0)--(Y0);
\draw[-{Latex[length=6.5pt]}] (Y0) to[bend right=19] (DELTAY);
\draw[->,shorten >= 1pt] (Y1)--(DELTAY);
\draw[-{Latex[length=6.5pt]}] (V0) to[bend left=20] (A1);
\draw[-{Latex[length=6.5pt]}] (V0) to[bend left=15] (Y1);
\draw[->,shorten >= 1pt] (S0)--(Y0);
\draw[-{Latex[length=6.5pt]}] (S0) to[bend left=15] (Y1);
\draw[->,shorten >= 1pt] (W0)--(Y0);
\draw[-{Latex[length=6.5pt]}] (W0) to[bend right=20] (A1);
\draw[->,shorten >= 1pt] (W0)--(W1);
\draw[-{Latex[length=6.5pt]}] (W1) to[bend right=17] (Y1);
\draw[->,shorten >= 1pt] (A1)--(W1);
\end{tikzpicture}
\caption{ }
\end{subfigure}
\hspace{2cm}
\begin{subfigure}{.4\textwidth}
\centering
\begin{tikzpicture}[>=triangle 45, font=\scriptsize]
\node[fill,circle,inner sep=0pt,minimum size=5pt,label={left:{$A_1$}}] (A1) at (0,0) {};
\node[fill,circle,inner sep=0pt,minimum size=5pt,label={right:{$\Delta Y$}}] (DELTAY) at (2,0) {};
\node[draw,circle,inner sep=0pt,minimum size=5pt,label={left:{$V_0$}}] (V0) at (-0.2,1) {};
\node[draw,circle,inner sep=0pt,minimum size=5pt,label={left:{$S_0$}}] (S0) at (1.8,1) {};
\node[fill,circle,inner sep=0pt,minimum size=5pt,label={below:{$W_0$}}] (W0) at (-0.25,-1) {};
\node[fill,circle,inner sep=0pt,minimum size=5pt,label={below:{$W_1$}}] (W1) at (1.5,-1) {};
\node[minimum size=5pt,label={left:{$a$}}] (a) at (1.25,0.15) {};
\node[minimum size=5pt,label={left:{$e-d$}}] (e-d) at (2.85,0.6) {};
\node[minimum size=5pt,label={left:{$b$}}] (b) at (0,0.6) {};
\node[minimum size=5pt,label={left:{$g$}}] (g) at (0,-0.6) {};
\node[minimum size=5pt,label={left:{$i$}}] (i) at (1.2,-0.35) {};
\node[minimum size=5pt,label={left, rotate=24:{$-f$}}] (-f) at (0.8,-0.6) {};
\node[minimum size=5pt,label={left:{$h$}}] (h) at (2.3,-0.7) {};
\node[minimum size=5pt,label={left:{$j$}}] (j) at (0.9,-1.2) {};
\draw[->,shorten >= 1pt] (A1)--(DELTAY);
\draw[->,shorten >= 1pt] (V0)--(A1);
\draw[->,shorten >= 1pt] (S0)--(DELTAY);
\draw[->,shorten >= 1pt] (W0)--(A1);
\draw[->,shorten >= 1pt] (W0)--(W1);
\draw[-{Latex[length=6.5pt]}] (W0) to[bend right=20] (DELTAY);
\draw[-{Latex[length=6.5pt]}] (W1) to[bend right=20] (DELTAY);
\draw[->,shorten >= 1pt] (A1)--(W1);
\end{tikzpicture}
\vspace{0.5cm}
\caption{ }
\end{subfigure}
\caption{Scenario in which the time-varying covariate $W_t$'s evolution depends on treatment and an auto-regressive process. \textbf{(a)} Natural diagram. \textbf{(b)} Compact representation. Illustrative linear model coefficients are shown as lowercase letters on each edge.}
\label{cdscenario3d}
\end{figure}
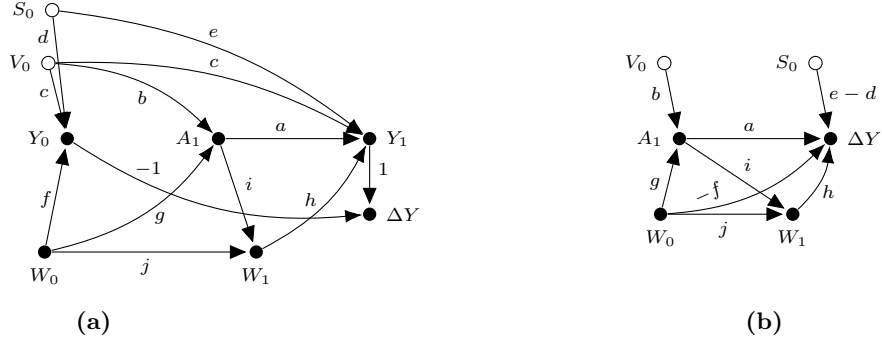

Figure~\ref{cdscenario3d} illustrates the situation in which treatment itself changes the covariate. From the compact representation in Panel (b), we confirm that conditioning on $\{W_0\}$ is minimally sufficient to block all backdoor paths. 

This scenario illustrates the case in which the post-treatment covariate must not be included in the adjustment set, as doing so would block a causal path from treatment to the outcome. In this particular scenario, adjusting for $W_1$ would recover only the direct effect rather than total effect, which is generally the estimand of interest.

In other cases, post-treatment covariates may act as colliders, and adjusting for them can open backdoor paths and introduce additional bias.

In conclusion, there is no universal rule governing adjustment for post-treatment covariates in \ac{DiD}. Across the four scenarios presented in this section, we have shown that such adjustment may be unnecessary but harmless, may serve as one among alternative sufficient strategies, may be the only option or may need to be avoided entirely. In each case, the appropriate decision follows from the structure of the causal model.

\subsection{Extensions to multiple time periods: time-varying treatment vs. staggered rollout}\label{ss:mult-time-periods}

The scenarios considered so far involve a single post-treatment period. Extending to multiple post-treatment periods requires distinguishing between two concepts: the treatment type and the rollout strategy. The treatment may be static, i.e., fixed (either one-time or sustained) or dynamic, i.e., it may change in response to evolving conditions. The rollout strategy may be simultaneous across all units or staggered, with different units starting treatment at different times. This distinction matters because these two concepts raise fundamentally different challenges for identification.

Static treatments are common in health policy and economics, where large units like hospitals, states or  countries implement one-time interventions whose effects persist over months or years. Dynamic treatments are more common in medicine and epidemiology, where individuals receive treatments that may change over days or weeks. Altering treatment in response to outcomes or patient factors may cause treatment-confounder feedback, in which a time-varying covariate mediates a prior treatment effect and confounds subsequent ones \cite{robinsNewApproachCausal1986,robinsMarginalStructuralModels2000}. Crucially, treatment-confounder feedback is a property of the type of treatment, not of the rollout strategy. It does not arise under static treatment regimes, even when implementation is staggered across units, because the treatment does not change with evolving conditions. In such cases, different cohorts simply begin the same static treatment at different times.

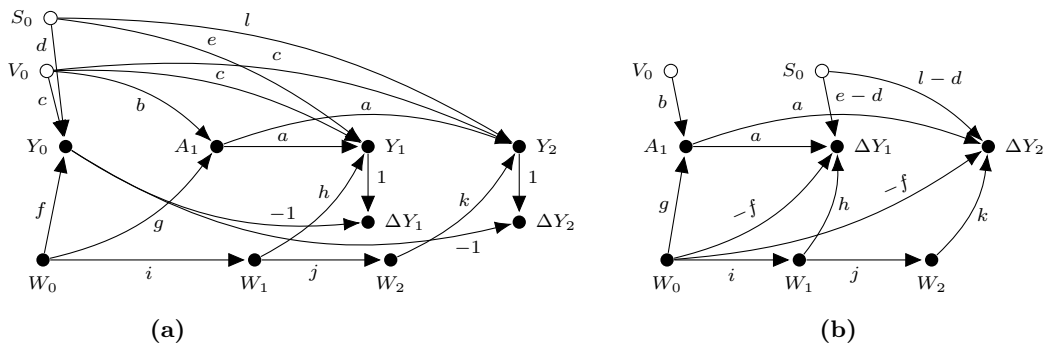
\begin{figure}[h!]
\begin{subfigure}{.4\textwidth}
\hspace{1cm}
\begin{tikzpicture}[>=triangle 45, font=\scriptsize]
\node[fill,circle,inner sep=0pt,minimum size=5pt,label={left:{$A_1$}}] (A1) at (0,0) {};
\node[fill,circle,inner sep=0pt,minimum size=5pt,label={left:{$Y_0$}}] (Y0) at (-2,0) {};
\node[fill,circle,inner sep=0pt,minimum size=5pt,label={right:{$Y_1$}}] (Y1) at (2,0) {};
\node[fill,circle,inner sep=0pt,minimum size=5pt,label={right:{$Y_2$}}] (Y2) at (4,0) {};
\node[draw,circle,inner sep=0pt,minimum size=5pt,label={left:{$V_0$}}] (V0) at (-2.25,1) {};
\node[draw,circle,inner sep=0pt,minimum size=5pt,label={left:{$S_0$}}] (S0) at (-2.2,1.7) {};
\node[fill,circle,inner sep=0pt,minimum size=5pt,label={right:{$\Delta Y_1$}}] (DELTAY1) at (2,-1) {};
\node[fill,circle,inner sep=0pt,minimum size=5pt,label={right:{$\Delta Y_2$}}] (DELTAY2) at (4,-1) {};
\node[fill,circle,inner sep=0pt,minimum size=5pt,label={below:{$W_0$}}] (W0) at (-2.3,-1.5) {};
\node[fill,circle,inner sep=0pt,minimum size=5pt,label={below:{$W_1$}}] (W1) at (0.5,-1.5) {};
\node[fill,circle,inner sep=0pt,minimum size=5pt,label={below:{$W_2$}}] (W2) at (2.3,-1.5) {};
\node[minimum size=5pt,label={left:{$e$}}] (e) at (0.25,1.4) {};
\node[minimum size=5pt,label={left:{$c$}}] (c) at (0.35,0.95) {};
\node[minimum size=5pt,label={left:{$c$}}] (c) at (-2,0.6) {};
\node[minimum size=5pt,label={left:{$c$}}] (c) at (1.1,1.2) {};
\node[minimum size=5pt,label={left:{$a$}}] (a) at (1.2,0.12) {};
\node[minimum size=5pt,label={left:{$d$}}] (d) at (-2,1.35) {};
\node[minimum size=5pt,label={left:{$b$}}] (b) at (-0.7,0.55) {};
\node[minimum size=5pt,label={left:{$1$}}] (1) at (2.5,-0.4) {};
\node[minimum size=5pt,label={left:{$-1$}}] (-1) at (1.3,-0.9) {};
\node[minimum size=5pt,label={left:{$1$}}] (1) at (4.5,-0.4) {};
\node[minimum size=5pt,label={left:{$-1$}}] (-1) at (3.75,-1.35) {};
\node[minimum size=5pt,label={left:{$i$}}] (i) at (-0.6,-1.67) {};
\node[minimum size=5pt,label={left:{$h$}}] (h) at (1.75,-0.6) {};
\node[minimum size=5pt,label={left:{$g$}}] (g) at (-0.45,-1.05) {};
\node[minimum size=5pt,label={left:{$f$}}] (f) at (-2,-0.85) {};
\node[minimum size=5pt,label={left:{$a$}}] (a) at (2.3,0.54) {};
\node[minimum size=5pt,label={left:{$l$}}] (l) at (0.7,1.68) {};
\node[minimum size=5pt,label={left:{$j$}}] (j) at (1.6,-1.67) {};
\node[minimum size=5pt,label={left:{$k$}}] (k) at (3.6,-0.7) {};
\draw[->,shorten >= 1pt] (A1)--(Y1);
\draw[-{Latex[length=6.5pt]}] (A1) to[bend left=20] (Y2);
\draw[->,shorten >= 1pt] (V0)--(Y0);
\draw[-{Latex[length=6.5pt]}] (Y0) to[bend right=19] (DELTAY1);
\draw[->,shorten >= 1pt] (Y1)--(DELTAY1);
\draw[-{Latex[length=6.5pt]}] (Y0) to[bend right=23] (DELTAY2);
\draw[->,shorten >= 1pt] (Y2)--(DELTAY2);
\draw[-{Latex[length=6.5pt]}] (V0) to[bend left=20] (A1);
\draw[-{Latex[length=6.5pt]}] (V0) to[bend left=15] (Y1);
\draw[-{Latex[length=6.5pt]}] (V0) to[bend left=15] (Y2);
\draw[->,shorten >= 1pt] (S0)--(Y0);
\draw[-{Latex[length=6.5pt]}] (S0) to[bend left=15] (Y1);
\draw[-{Latex[length=6.5pt]}] (S0) to[bend left=15] (Y2);
\draw[->,shorten >= 1pt] (W0)--(Y0);
\draw[-{Latex[length=6.5pt]}] (W0) to[bend right=20] (A1);
\draw[->,shorten >= 1pt] (W0)--(W1);
\draw[->,shorten >= 1pt] (W1)--(W2);
\draw[-{Latex[length=6.5pt]}] (W1) to[bend right=17] (Y1);
\draw[-{Latex[length=6.5pt]}] (W2) to[bend right=17] (Y2);
\end{tikzpicture}
\caption{ }
\end{subfigure}
\hspace{2cm}
\begin{subfigure}{.4\textwidth}
\centering
\begin{tikzpicture}[>=triangle 45, font=\scriptsize]
\node[fill,circle,inner sep=0pt,minimum size=5pt,label={left:{$A_1$}}] (A1) at (0,0) {};
\node[fill,circle,inner sep=0pt,minimum size=5pt,label={right:{$\Delta Y_1$}}] (DELTAY1) at (2,0) {};
\node[fill,circle,inner sep=0pt,minimum size=5pt,label={right:{$\Delta Y_2$}}] (DELTAY2) at (4,0) {};
\node[draw,circle,inner sep=0pt,minimum size=5pt,label={left:{$V_0$}}] (V0) at (-0.2,1) {};
\node[draw,circle,inner sep=0pt,minimum size=5pt,label={left:{$S_0$}}] (S0) at (1.8,1) {};
\node[fill,circle,inner sep=0pt,minimum size=5pt,label={below:{$W_0$}}] (W0) at (-0.25,-1.5) {};
\node[fill,circle,inner sep=0pt,minimum size=5pt,label={below:{$W_1$}}] (W1) at (1.5,-1.5) {};
\node[fill,circle,inner sep=0pt,minimum size=5pt,label={below:{$W_2$}}] (W2) at (3.25,-1.5) {};
\node[minimum size=5pt,label={left:{$a$}}] (a) at (1.25,0.12) {};
\node[minimum size=5pt,label={left:{$a$}}] (a) at (1.8,0.53) {};
\node[minimum size=5pt,label={left:{$e-d$}}] (e-d) at (2.85,0.68) {};
\node[minimum size=5pt,label={left:{$l-d$}}] (l-d) at (3.9,0.9) {};
\node[minimum size=5pt,label={left:{$b$}}] (b) at (0,0.6) {};
\node[minimum size=5pt,label={left:{$g$}}] (g) at (0.03,-0.8) {};
\node[minimum size=5pt,label={left, rotate=24:{$-f$}}] (-f) at (1.2,-0.73) {};
\node[minimum size=5pt,label={left, rotate=24:{$-f$}}] (-f) at (3.2,-0.4) {};
\node[minimum size=5pt,label={left:{$h$}}] (h) at (2.42,-0.75) {};
\node[minimum size=5pt,label={left:{$k$}}] (k) at (4.25,-0.9) {};
\node[minimum size=5pt,label={left:{$i$}}] (i) at (0.9,-1.7) {};
\node[minimum size=5pt,label={left:{$j$}}] (j) at (2.55,-1.7) {};
\draw[->,shorten >= 1pt] (A1)--(DELTAY1);
\draw[->,shorten >= 1pt] (V0)--(A1);
\draw[->,shorten >= 1pt] (S0)--(DELTAY1);
\draw[->,shorten >= 1pt] (W0)--(A1);
\draw[->,shorten >= 1pt] (W0)--(W1);
\draw[->,shorten >= 1pt] (W1)--(W2);
\draw[-{Latex[length=6.5pt]}] (W0) to[bend right=20] (DELTAY1);
\draw[-{Latex[length=6.5pt]}] (W0) to[bend right=15] (DELTAY2);
\draw[-{Latex[length=6.5pt]}] (W1) to[bend right=20] (DELTAY1);
\draw[-{Latex[length=6.5pt]}] (W2) to[bend right=20] (DELTAY2);
\draw[-{Latex[length=6.5pt]}] (S0) to[bend left=20] (DELTAY2);
\draw[-{Latex[length=6.5pt]}] (A1) to[bend left=20] (DELTAY2);
\end{tikzpicture}
\caption{ }
\end{subfigure}
\caption{Scenario in which a static treatment may have time-varying effects. \textbf{(a)} Natural diagram. \textbf{(b)} Compact representation. Illustrative linear model coefficients are shown as lowercase letters on each edge.}
\label{cdscenario4}
\end{figure}

We illustrate these points with two scenarios. Figure \ref{cdscenario4} illustrates a setting with two post-treatment periods ($t = 2$) and a static treatment, $A_1$, representing a one-time intervention with potentially persistent effects on subsequent outcomes.

From the compact representation in Panel (b), we see that $\{W_0\}$ is a minimally sufficient adjustment set to identify the effect of treatment on both $\Delta Y_1$ and $\Delta Y_2$. Of note, across all scenarios in this section, we hold the treatment effect constant across post-treatment periods. This design choice isolates each estimator's handling of confounding through adjustment set specification as the sole source of potential bias, thereby enabling a fair comparison using identical estimator implementations.

These principles extend naturally to more complex multi-period settings, including staggered rollout. In such cases, we recommend constructing a separate causal diagram for each treatment implementation time, as this allows treatment assignment mechanisms to differ across cohorts. For example, early-implementing states of paid family leave policies such as California (2002) and New Jersey (2008) were influenced by strong labor coalitions and pre-existing disability insurance systems \cite{milkmanUnfinishedBusinessPaid2013}. Later-implementing states without such infrastructure faced fundamentally different institutional and political conditions \cite{glynnNATIONALACADEMYSOCIAL}. Thus, constructing distinct causal diagrams for each cohort encourages careful consideration of how treatment assignment mechanisms evolve over time.

\begin{figure}[h!]
\begin{subfigure}{.4\textwidth}
\begin{tikzpicture}[>=triangle 45, font=\scriptsize]
\node[fill,circle,inner sep=0pt,minimum size=5pt,label={left:{$A_1$}}] (A1) at (0,0) {};
\node[fill,circle,inner sep=0pt,minimum size=5pt, label={[xshift=1mm,yshift=-0.5mm]{$A_2$}}] (A2) at (3.5,0) {};
\node[fill,circle,inner sep=0pt,minimum size=5pt,label={left:{$Y_0$}}] (Y0) at (-2,0) {};
\node[fill,circle,inner sep=0pt,minimum size=5pt,label={right:{$Y_1$}}] (Y1) at (2,0) {};
\node[fill,circle,inner sep=0pt,minimum size=5pt,label={right:{$Y_2$}}] (Y2) at (5.5,0) {};
\node[draw,circle,inner sep=0pt,minimum size=5pt,label={left:{$V_0$}}] (V0) at (-2.25,1) {};
\node[draw,circle,inner sep=0pt,minimum size=5pt,label={left:{$S_0$}}] (S0) at (-2.2,1.7) {};
\node[fill,circle,inner sep=0pt,minimum size=5pt,label={below:{$\Delta Y_1$}}] (DELTAY1) at (2,-1) {};
\node[fill,circle,inner sep=0pt,minimum size=5pt,label={below:{$\Delta Y_2$}}] (DELTAY2) at (5.5,-1) {};
\node[fill,circle,inner sep=0pt,minimum size=5pt,label={below:{$W_0$}}] (W0) at (-2.3,-2) {};
\node[fill,circle,inner sep=0pt,minimum size=5pt,label={below:{$W_1$}}] (W1) at (0.5,-2) {};
\node[fill,circle,inner sep=0pt,minimum size=5pt,label={below:{$W_2$}}] (W2) at (4,-2) {};
\node[minimum size=5pt,label={left:{$e$}}] (e) at (0.25,1.4) {};
\node[minimum size=5pt,label={left:{$c$}}] (c) at (0.35,0.95) {};
\node[minimum size=5pt,label={left:{$c$}}] (c) at (-2,0.6) {};
\node[minimum size=5pt,label={left:{$c$}}] (c) at (1.4,1.36) {};
\node[minimum size=5pt,label={left:{$a$}}] (a) at (1.27,0.12) {};
\node[minimum size=5pt,label={left:{$d$}}] (d) at (-2,1.35) {};
\node[minimum size=5pt,label={left:{$b$}}] (b) at (-0.7,0.55) {};
\node[minimum size=5pt,label={left:{$1$}}] (1) at (2.45,-0.4) {};
\node[minimum size=5pt,label={left:{$-1$}}] (-1) at (1.3,-0.9) {};
\node[minimum size=5pt,label={left:{$1$}}] (1) at (5.95,-0.4) {};
\node[minimum size=5pt,label={left:{$-1$}}] (-1) at (3.5,-1.75) {};
\node[minimum size=5pt,label={left:{$j$}}] (j) at (-0.6,-2.18) {};
\node[minimum size=5pt,label={left:{$i$}}] (i) at (1.78,-0.8) {};
\node[minimum size=5pt,label={left:{$g$}}] (g) at (-0.6,-1.05) {};
\node[minimum size=5pt,label={left:{$h$}}] (h) at (0.65,-0.7) {};
\node[minimum size=5pt,label={left:{$f$}}] (f) at (-1.95,-0.95) {};
\node[minimum size=5pt,label={left:{$a$}}] (a) at (4.65,0.12) {};
\node[minimum size=5pt,label={left:{$q$}}] (q) at (1.5,1.75) {};
\node[minimum size=5pt,label={left:{$k$}}] (k) at (2.5,-2.18) {};
\node[minimum size=5pt,label={left:{$l$}}] (l) at (3.2,-1.2) {};
\node[minimum size=5pt,label={left:{$n$}}] (n) at (5.23,-0.9) {};
\node[minimum size=5pt,label={left:{$m$}}] (m) at (4.3,-0.9) {};
\node[minimum size=5pt,label={left:{$p$}}] (p) at (2.3,0.5) {};
\node[minimum size=5pt,label={left:{$b$}}] (b) at (3,0.55) {};
\draw[->,shorten >= 1pt] (A1)--(Y1);
\draw[->,shorten >= 1pt] (A2)--(Y2);
\draw[-{Latex[length=6.5pt]}] (A1) to[bend left=20] (A2);
\draw[->,shorten >= 1pt] (V0)--(Y0);
\draw[-{Latex[length=6.5pt]}] (Y0) to[bend right=19] (DELTAY1);
\draw[->,shorten >= 1pt] (Y1)--(DELTAY1);
\draw[-{Latex[length=6.5pt]}] (Y0) to[bend right=26] (DELTAY2);
\draw[->,shorten >= 1pt] (Y2)--(DELTAY2);
\draw[-{Latex[length=6.5pt]}] (V0) to[bend left=20] (A1);
\draw[-{Latex[length=6.5pt]}] (V0) to[bend left=16] (A2);
\draw[-{Latex[length=6.5pt]}] (V0) to[bend left=15] (Y1);
\draw[-{Latex[length=6.5pt]}] (V0) to[bend left=15] (Y2);
\draw[->,shorten >= 1pt] (S0)--(Y0);
\draw[-{Latex[length=6.5pt]}] (S0) to[bend left=15] (Y1);
\draw[-{Latex[length=6.5pt]}] (S0) to[bend left=15] (Y2);
\draw[->,shorten >= 1pt] (W0)--(Y0);
\draw[-{Latex[length=6.5pt]}] (W0) to[bend right=20] (A1);
\draw[->,shorten >= 1pt] (W0)--(W1);
\draw[->,shorten >= 1pt] (W1)--(W2);
\draw[-{Latex[length=6.5pt]}] (W1) to[bend right=17] (Y1);
\draw[-{Latex[length=6.5pt]}] (W1) to[bend right=30] (A2);
\draw[->,shorten >= 1pt] (A1)--(W1);
\draw[->,shorten >= 1pt] (A2)--(W2);
\draw[-{Latex[length=6.5pt]}] (W2) to[bend right=17] (Y2);
\end{tikzpicture}
\caption{ }
\end{subfigure}
\hspace{3cm}
\begin{subfigure}{.4\textwidth}
\centering
\begin{tikzpicture}[>=triangle 45, font=\scriptsize]
\node[fill,circle,inner sep=0pt,minimum size=5pt,label={left:{$A_1$}}] (A1) at (0,0) {};
\node[fill,circle,inner sep=0pt,minimum size=5pt,label={right:{$\Delta Y_1$}}] (DELTAY1) at (2,0) {};
\node[fill,circle,inner sep=0pt,minimum size=5pt,label={above:{$A_2$}}] (A2) at (4,0) {};
\node[fill,circle,inner sep=0pt,minimum size=5pt,label={right:{$\Delta Y_2$}}] (DELTAY2) at (6,0) {};
\node[draw,circle,inner sep=0pt,minimum size=5pt,label={left:{$V_0$}}] (V0) at (-0.2,1) {};
\node[draw,circle,inner sep=0pt,minimum size=5pt,label={left:{$S_0$}}] (S0) at (1.8,1.5) {};
\node[fill,circle,inner sep=0pt,minimum size=5pt,label={below:{$W_0$}}] (W0) at (-0.25,-2) {};
\node[fill,circle,inner sep=0pt,minimum size=5pt,label={below:{$W_1$}}] (W1) at (1.1,-2) {};
\node[fill,circle,inner sep=0pt,minimum size=5pt,label={below:{$W_2$}}] (W2) at (4.75,-2) {};
\node[minimum size=5pt,label={left:{$a$}}] (a) at (1.27,0.12) {};
\node[minimum size=5pt,label={left:{$p$}}] (p) at (2.48,0.6) {};
\node[minimum size=5pt,label={left:{$a$}}] (a) at (5.15,0.12) {};
\node[minimum size=5pt,label={left:{$e-d$}}] (e-d) at (2.85,1.1) {};
\node[minimum size=5pt,label={left:{$q-d$}}] (q-d) at (4.75,1.3) {};
\node[minimum size=5pt,label={left:{$b$}}] (b) at (0,0.6) {};
\node[minimum size=5pt,label={left:{$b$}}] (b) at (0.7,1.2) {};
\node[minimum size=5pt,label={left:{$g$}}] (g) at (0.05,-1.1) {};
\node[minimum size=5pt,label={left, rotate=24:{$-f$}}] (-f) at (1.7,-0.55) {};
\node[minimum size=5pt,label={left, rotate=24:{$-f$}}] (-f) at (5.5,-0.34) {};
\node[minimum size=5pt,label={left:{$h$}}] (h) at (0.93,-0.75) {};
\node[minimum size=5pt,label={left:{$m$}}] (m) at (4.95,-0.9) {};
\node[minimum size=5pt,label={left:{$n$}}] (n) at (6.1,-1.15) {};
\node[minimum size=5pt,label={left:{$j$}}] (j) at (0.7,-2.15) {};
\node[minimum size=5pt,label={left:{$k$}}] (k) at (3.1,-2.15) {};
\node[minimum size=5pt,label={left:{$l$}}] (l) at (3.5,-1.2) {};
\node[minimum size=5pt,label={left:{$i$}}] (i) at (2.35,-0.85) {};
\draw[->,shorten >= 1pt] (A1)--(DELTAY1);
\draw[->,shorten >= 1pt] (V0)--(A1);
\draw[->,shorten >= 1pt] (S0)--(DELTAY1);
\draw[->,shorten >= 1pt] (W0)--(A1);
\draw[->,shorten >= 1pt] (W0)--(W1);
\draw[->,shorten >= 1pt] (W1)--(W2);
\draw[->,shorten >= 1pt] (A1)--(W1);
\draw[->,shorten >= 1pt] (A2)--(W2);
\draw[-{Latex[length=6.5pt]}] (W0) to[bend right=20] (DELTAY1);
\draw[-{Latex[length=6.5pt]}] (W1) to[bend right=25] (DELTAY1);
\draw[-{Latex[length=6.5pt]}] (W0) to[bend right=-3] (DELTAY2);
\draw[-{Latex[length=6.5pt]}] (W1) to[bend right=20] (A2);
\draw[-{Latex[length=6.5pt]}] (W2) to[bend right=20] (DELTAY2);
\draw[-{Latex[length=6.5pt]}] (S0) to[bend left=14] (DELTAY2);
\draw[-{Latex[length=6.5pt]}] (A1) to[bend left=19] (A2);
\draw[-{Latex[length=6.5pt]}] (V0) to[bend left=13] (A2);
\draw[->,shorten >= 1pt] (A2)--(DELTAY2);
\end{tikzpicture}
\caption{ }
\end{subfigure}
\caption{Scenario with a dynamic treatment regime and treatment-confounder feedback via $W_1$. \textbf{(a)} Natural diagram. \textbf{(b)} Compact representation. Illustrative linear model coefficients are shown as lowercase letters on each edge.}
\label{cdscenario4_2}
\end{figure}
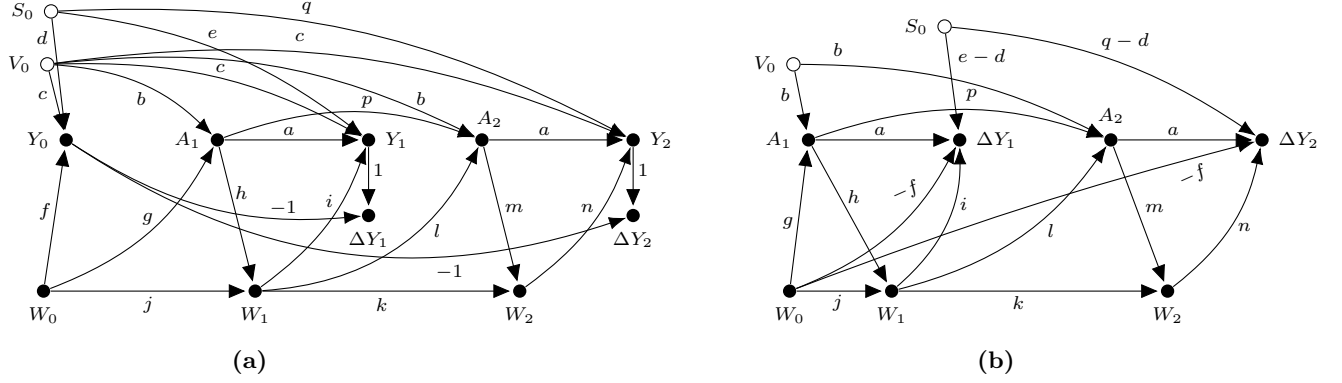

Figure \ref{cdscenario4_2} illustrates a dynamic treatment regime with two treatment times ($A_1$ and $A_2$) in which $W_1$ is both a mediator of the effect of $A_1$ on $Y_1$ and a confounder of the effect of $A_2$ on $Y_2$. Of note, we evaluate treatment effect estimates at each time period separately, rather than examining the entire treatment trajectory. 

Examining Panel (b) shows that $\{W_0\}$ is a minimally sufficient adjustment set for identifying the effect of $A_1$ on $\Delta Y_1$, whereas $\{W_0,W_1\}$ is minimally sufficient for the effect of $A_2$ on $\Delta Y_2$.

The key takeaway from these two scenarios is that the distinction between the type of treatment and the rollout strategy has direct consequences for covariate adjustment: treatment-confounder feedback, which requires adjusting for covariates that are simultaneously mediators and confounders, arises only under dynamic treatments, not from staggered implementation alone.

\section{Estimation}\label{SecEst}

We have shown that causal diagrams coupled with graphical criteria can guide the selection of adjustment sets for \ac{DiD} identification. We now turn to the question of how popular estimation procedures implement the required covariate adjustment. For each estimator, we introduce a compact label that highlights how covariates enter the procedure, which can be through outcome modeling, weighting, or both. These labels are intended to make transparent the way each estimator incorporates covariates, since, as we will show, different estimation procedures may default to different adjustment sets even when targeting the same sufficient adjustment set. Below and in Table~\ref{tab:estimator_summaries}, we briefly describe each estimator (see Appendix for more details).

We begin with estimators that implement covariate adjustment through outcome modeling. The $\Delta Y$ method, denoted $\Delta Y(X)$, regresses outcome changes on covariates and a treatment group indicator, taking the coefficient on the treatment indicator as the estimate. The standard \ac{TWFE} model, denoted $Y(X)$, regresses the outcome on covariates, treatment and time indicators, and a treatment-time interaction. The coefficient on this interaction term is taken as the treatment effect estimate. De Chaisemartin and D'Haultfœuille's estimator \cite{dechaisemartinTwoWayFixedEffects2020}, denoted $\epsilon( \Delta Y(\Delta X))$, regresses outcome changes on covariate changes among control units; computes residuals, $\epsilon$, between observed and predicted outcome changes for all units; then regresses those residuals on the treatment indicator to obtain the treatment effect estimate. Unique among the estimators we consider, this estimator adjusts for covariate \emph{changes}. Heckman's outcome regression \cite{heckmanMatchingEconometricEvaluation1997a,heckmanCharacterizingSelectionBias1998}, denoted $\epsilon(\Delta Y(X))$, also fits an outcome model on control units, but regresses outcome changes on covariates levels rather than covariate changes. The fitted model is then used to predict counterfactual outcomes for treated units, and the treatment effect estimate is the average difference between observed and predicted outcomes among treated units.

\begin{table}[ht]
    \centering
    \caption{Summary of the estimators}
    \begin{tabular}{l|ccrcc}
        Name & Propensity scores & \multicolumn{2}{c}{Weights} & Outcome predictions & Estimator \\
        \hline
        $\Delta Y (X)$  &  &  &  &  & $\theta$ from \\
                    &  &  &  &  & $\Delta Y \sim \theta A + \bm \beta'\mathbf{X}$\\
        \hline
        $Y(X)$        &  &  &  &  & $\theta$ from \\
        TWFE            &  &  &  &  & $Y_t \sim \ldots + \theta AP + \bm \beta_t'\mathbf{X}$ \\
        \hline
        $\epsilon(\Delta Y(\Delta X))$ &  & &  & $\widehat{\Delta Y}$ from                  & $\mathbb{E}_{trt}\left[\Delta Y-\widehat{\Delta Y}\right]-$ \\
        de C-D'H    &  & &  & $\Delta Y \sim \bm {\beta'} \Delta \mathbf{X} | A=0$    & $\mathbb{E}_{ctrl}\left[\Delta Y-\widehat{\Delta Y}\right]$ \\
        \hline
        $\epsilon(\Delta Y(X))$ &  &  & & $\widehat{\Delta Y}$ from           & $\mathbb{E}_{trt}\left[\Delta Y-\widehat{\Delta Y}\right]$ \\
        Heckman      &  &  & & $\Delta Y \sim \bm \beta'\mathbf{X} | A=0$    &  \\
        \hline
        $w(X) \Delta Y$ & $e(\mathbf{X})$ from          & $1^{\dagger}$ & $A = 1$ &  
                        & $\mathbb{E}_{trt}\left[w\Delta Y\right]-$ \\
        Abadie IPW            & $A \sim \bm \pi' \mathbf{X}$  & $\dfrac{e(\mathbf{X})}{1 - e(\mathbf{X})}^{\dagger}$ & $A = 0$ & 
                        & $\mathbb{E}_{ctrl}\left[w\Delta Y\right]$ \\
        \hline
        $w(X)\epsilon(\Delta Y(X)) $    & $e(\mathbf{X})$ from          & $1^{\dagger}$ & $A = 1$        
                                & $\widehat{\Delta Y}$ from & 
                                $\mathbb{E}_{trt}\left[w\left(\Delta Y-\widehat{\Delta Y}\right)\right]-$\\
        S'A-Z DR     & $A \sim \bm\pi' \mathbf{X}$   & $\dfrac{e(\mathbf{X})}{1 - e(\mathbf{X})}^{\dagger} $ & $ A = 0$ 
                                & $\Delta Y \sim \bm \beta'\mathbf{X} | A=0$ & 
                                $\mathbb{E}_{ctrl}\left[w\left(\Delta Y-\widehat{\Delta Y}\right)\right]$ \\
        \hline        
        $w_g(X)Y$ & $e_{g}(\mathbf{X})$ from        & $\dfrac{e_1}{e_{G}(\mathbf{X})}$ & $A=1$  & & $\theta$ from$^{\ddagger}$ \\
        Stuart g-PS            & $G \sim \bm \pi_g \mathbf{X}$ & $\dfrac{e_1}{e_{G}(\mathbf{X})}$ & $A=0$  & & $Y_{t} \sim \ldots+ \theta AP $ \\
        \hline
        $w_t(X)Y$  & $e_t(\mathbf{X})$ from        & $\dfrac{1}{e_t(\mathbf{X})}$ & $A = 1$ &       & $\theta$ from$^{\ddagger}$  \\
         Stuart t-PS           & $A\sim \bm \pi'_t \mathbf{X}$ & $\dfrac{1}{1 - e_t(\mathbf{X})}$ & $A = 0$ &   & $Y_{t} \sim \ldots + \theta AP$ \\
        \hline
        $w_t^{ATT}(X) \Delta Y$   & $e_t(\mathbf{X})$ from          & $1$ & $A_t = 1$ &  &  $\mathbb{E}_{trt}\left[w_t\Delta Y\right]-$ \\
        Myint ATT          & $A_t \sim \bm \pi_t'\mathbf{X}$  & $\dfrac{e_t(\mathbf{X})}{1-e_t(\mathbf{X})}$ & $A_t = 0$ &  & $\mathbb{E}_{ctrl}\left[w_t\Delta Y\right]$\\
        \hline
        $w_t^{ATE}(X) \Delta Y$   & $e_t(\mathbf{X})$ from  & $\dfrac{p^{trt}}{e_t(\mathbf{X})}$ & $A_t = 1$ & & $\mathbb{E}_{trt}\left[w_t\Delta Y\right]-$ \\
        Myint ATE            & $A_t\sim \bm \pi_t'\mathbf{X}$    & $\dfrac{1-p^{trt}}{1-e_t(\mathbf{X})}$ & $A_t = 0$ &  & $\mathbb{E}_{ctrl}\left[w_t\Delta Y\right]$\\

        \hline
        \multicolumn{6}{p{0.8\textwidth}}{ $^{\dagger}$Normalized to sum to one within each group. \newline
        $^{\ddagger}$Weighted regression.\newline
        The omitted regression terms indicated by $\ldots$ are $\eta_0 + \eta_A A + \eta_P P$, where $P=I\{t=1\}$ 
        }
    \end{tabular}
    \label{tab:estimator_summaries}
    
\end{table}

Next, we consider an estimator based on weighting. Abadie's \ac{IPW} estimator \cite{abadieSemiparametricDifferenceindifferencesEstimators2005}, denoted $w(X) \Delta Y$, estimates a propensity score for treatment assignment given covariates from both treated and control units and uses inverse probability weights to weight the comparison group toward the distribution of the treated group. The treatment effect contrasts outcome changes in the treated group with those in the reweighted comparison group. Sant'Anna and Zhao's doubly robust estimator \cite{santannaDoublyRobustDifferenceindifferences2020a}, denoted $w(X) \epsilon(\Delta Y(X))$, combines outcome modeling with inverse probability weighting, thus protecting against misspecification of either the outcome or propensity score model. 

A brief note about data structure. If we have panel data, the $\Delta Y$ in the Heckman's outcome regression, Abadie's \ac{IPW} and Sant'Anna and Zhao's doubly robust estimators can be computed directly for each unit. With repeated cross-sections, these estimators operate on outcomes at each time point separately.

Finally, we consider estimators that combine weighting with an outcome procedure in a two-step fashion. Unlike the doubly robust approach, these estimators apply weighting and outcome modeling sequentially rather than jointly, and do not offer protection against misspecification of either model alone. Stuart et al. \cite{stuartUsingPropensityScores2014a} first estimate propensity score weights and then apply them to a \ac{TWFE} regression. In our implementation, we omit covariates from the \ac{TWFE} regression, implementing covariate adjustment entirely through the weights, though Stuart et al.~also consider specifications that include covariates in both steps. The authors propose two variants: a multinomial version that fits a single propensity score model across all four treatment-by-period groups, weighting each toward a reference group, denoted $w_g(X)Y$, and a time-specific version that fits separate binary propensity score models at each time period, denoted $w_t(X)Y$. Myint \cite{myintControllingTimevaryingConfounding2023} develops sequential \ac{IPW} estimators for settings with potential time-invariant and time-varying confounding. The original method constructs cumulative weights as the product of propensity scores across time points, targeting the treatment effect over an entire treatment regime. Since our goal is to estimate the treatment effect at each time point separately, we do not use cumulative weights. Instead, we fit propensity score models at each time point, where the adjustment set may include covariates from current and prior periods. These weights are then applied in two ways. In the first version, denoted $w_t^{ATT}(X) \Delta Y$, \ac{ATT} weights are constructed and the weighted outcome changes are regressed on the treatment indicator, with no additional covariates. In the second version, denoted $w_t^{ATE}(X) \Delta Y$, \ac{ATE} weights are constructed and combined with the Callaway and Sant'Anna \ac{DiD} framework \cite{callawayDidTreatmentEffects2022}. Since we do not include covariates at this stage, this second step reduces to a difference in weighted mean outcome changes between treated and control groups.

\subsection{Changing default covariate handling}\label{ss:adjustment_sets}

Each estimator handles covariates differently by default, depending on the data format it requires (wide, long, or differenced) and on how it incorporates the supplied covariates into the estimation procedure. As a result, different estimators may default to different adjustment sets even when the goal is to adjust for the same confounders. Some estimators handle this naturally, while others require additional strategies to capture a sufficient adjustment set, as we describe below.

The $\Delta Y$ method ($\Delta Y (X)$) and the first step of both Myint procedures ($w_t^{ATT}(X) \Delta Y$ and $w_t^{ATE}(X) \Delta Y$) operate on wide-format data, where each covariate at each time point appears as a separate column. Thus, all supplied covariates (time-invariant or time-varying) contribute to the effective adjustment set directly, and no modification to the default covariate handling is needed.

For the remaining estimators, which operate on long-format or differenced data, the default covariate handling may not match a sufficient adjustment set. The standard \ac{TWFE} model ($Y(X)$) assigns a single coefficient to each covariate, pooled across time periods. For time-invariant covariates, their contribution appears identically at both time periods and cancels from the implied difference, leaving the treatment effect estimate unchanged, even when the covariate is part of a sufficient adjustment set. For time-varying covariates, the values differ across periods, so the contribution does not cancel entirely. However, with a single pooled coefficient, it is ambiguous whether the estimator effectively adjusts for $X_0$, $X_1$ or some combination of both. To address the first limitation, we specify an augmented \ac{TWFE} model with covariate-by-time interactions, which assigns time-specific coefficients, thus preventing time-invariant covariates from canceling out. However, when it comes to time-varying covariates, this augmented specification produces a perhaps unexpected effective adjustment set: the model adjusts for $X_0$ at $t=0$ and $X_1$ at $t=1$ separately, rather than for both covariates at each time period, which may or may not constitute a sufficient adjustment set. To address this second potential limitation, we create time-constant copies of time-varying covariates, i.e., separate variables that fix the covariate at its value from a specific period and hold it constant across all periods in the long-data format (e.g., $X^{dup}_0$ takes the value of $X_0$ at both $t=0$ and $t=1$).

De Chaisemartin and D'Haultfœuille's estimator ($\epsilon(\Delta Y(\Delta X))$) operates on covariate changes, which produces a similar problem to the standard \ac{TWFE} model: time-invariant covariates difference out to zero and therefore do not contribute to the treatment effect estimate, while time-varying covariates give rise to an ambiguous effective adjustment set. To address this, we interact time-invariant covariates and time-constant copies of time-varying covariates with time directly in the dataset prior to estimation. Differencing these interaction terms recovers the original covariate levels, e.g., $\Delta (X_0 \times t)=X_0$ and $\Delta (X^{dup}_0 \times t)=X^{dup}_0$ and $\Delta (X^{dup}_1 \times t)=X^{dup}_1$, thus preserving these covariates in the estimation procedure. We implement this estimator via the \texttt{did\_multipleft} function from the \texttt{DIDmultiplegt} package \cite{quispeDIDmltiplegtpackage}.

Heckman's outcome regression ($\epsilon(\Delta Y (X))$), Abadie's \ac{IPW} ($w(X)\Delta Y$), and Sant'Anna and Zhao's doubly robust ($w(X)\epsilon(\Delta Y (X))$) estimators share the same default covariate handling because we implement all three via the \texttt{att\_gt} function in the \texttt{did} package \cite{callawayDidTreatmentEffects2022}. The package's default handling of time-varying covariates is to use only the baseline value; for instance, when time-varying $X_t$ is given, only $X_0$ enters the model. To include both baseline and post-treatment covariate values in the effective adjustment set, we replace the time-varying covariates with their time-constant copies (e.g., $X_0^{dup}$ and $X_1^{dup}$ to adjust for both $X_0$ and $X_1$). We also consider an alternative (but equivalent) adjustment set consisting of the baseline covariate and the covariate change between time periods, as proposed by Caetano and Callaway \cite{caetanoDifferenceDifferencesTimeVarying2024a,caetanoDifferenceinDifferencesWhenParallel2024}. Since $\{X_0, X_1-X_0\}$ is a linear transformation of $\{X_0, X_1\}$, the two sets span the same information. In scenarios where $\{X_0, X_1\}$ constitutes a sufficient adjustment set, $\{X_0, X_1-X_0\}$ is equally sufficient.

Finally, Stuart et al.'s procedures each handle covariates differently in the weighting step. In the group-specific version ($w_g(X)Y$), the propensity score model is fitted on data pooled across treatment-by-time groups, making the effective adjustment set difficult to characterize. In the time-specific version ($w_t(X)Y$), separate propensity score models are fitted at each time period, so each model adjusts only for the covariate values observed at that time. To override these defaults, we again supply time-constant copies of time-varying covariates, ensuring that each propensity score model can condition on covariate values from any time period.

These strategies ensure that, for each estimator, we can construct effective adjustment sets that align with the sufficient adjustment sets for identification.

\section{Simulation results}\label{SecRes}

We simulated each scenario described in Sections \ref{ss:time-invariant-only}-\ref{ss:mult-time-periods} and applied all estimators introduced above, with one exception: in Scenario \ref{ss:mult-time-periods} (2), which features treatment-confounder feedback, we only considered the $\Delta Y$ method. Specifically, we estimate the time-specific effects of $A_1$ and $A_2$ via separate models, one for each time point. We selected this procedure because of its simplicity and the fact that it operates on covariates in wide format, which allows us to specify the adjustment set transparently at each time period. While more recent methods \cite{rensonIdentifyingEstimatingEffects2023,shahnStructuralNestedMean2022} offer stronger theoretical guarantees and improved efficiency in this context, our objective is to illustrate how bias depends on the adjustment set, for which a simple approach suffices. We used a sample size of 5000 and 500 repetitions throughout. The code to reproduce all results is available at \url{https://github.com/DS-Rodrigues/Variable_Selection_DID}.

\begin{landscape}
\begin{figure}
    \centering
    \includegraphics[width=1\linewidth]{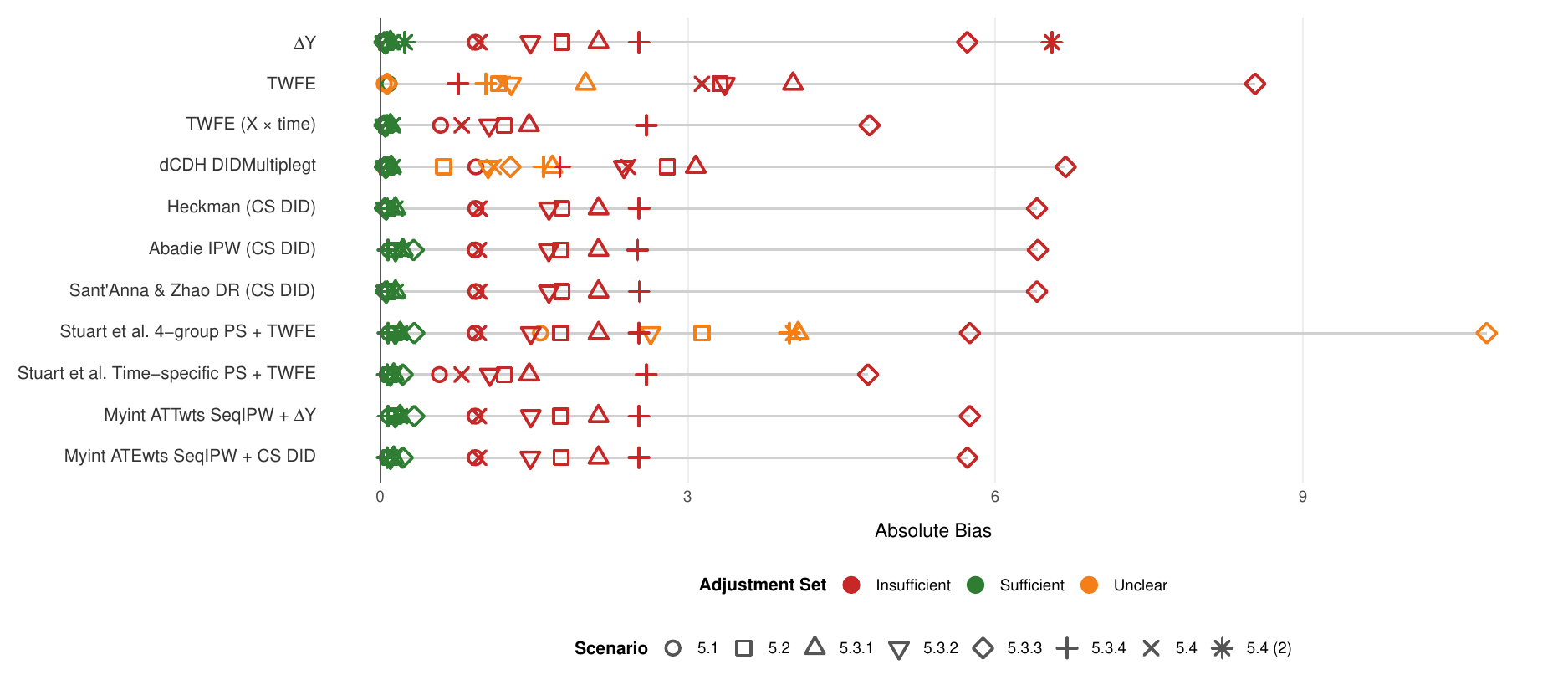}
    \caption{Mean absolute bias for each estimator, averaged across adjustment set categories (i.e., sufficient, insufficient and unclear) and replications in each scenario.}
    \label{fig:final_results}
\end{figure}
\end{landscape}

Figure~\ref{fig:final_results} shows the absolute bias of each estimator across all scenarios, stratified by adjustment set category: sufficient, insufficient or unclear. We classify an adjustment set as sufficient if it satisfies the backdoor criterion, insufficient if it does not, and unclear if the effective adjustment set cannot be determined a priori from the mechanics of the estimator. The simulation results confirm our theoretical findings. Across all scenarios, sufficient adjustment sets consistently yield low bias, with outcome regression approaches performing best and \ac{IPW}-based approaches showing slightly larger but still negligible bias. Insufficient adjustment sets produce substantial bias across all estimators, the magnitude of which reflects the strength of confounding in the underlying data-generating process. For unclear adjustment sets, the bias ranges from near zero to the largest value observed in the study. This finding highlights a practical concern: when the effective adjustment set cannot be clearly characterized, the identification conditions cannot be verified.

To illustrate these patterns concretely, we highlight aspects of selected scenarios that best demonstrate the role of the adjustment set. Full results are provided in the Supplementary Material.

In Scenario \ref{ss:time-invariant-only}, unconditional parallel trends hold, and conditioning on $Z_1$ violates this assumption. Consequently, as expected, estimators that adjust for $Z_1$ yield biased estimates. Although adjusting for $Z_1$ in isolation may appear unlikely in practice, several estimators do so implicitly. For instance, the augmented \ac{TWFE} model with $Z$ interacted with time adjusts for $Z_0$ at $t=0$ and $Z_1$ at $t=1$ separately, producing bias of approximately $0.3$. Stuart et al.'s time-specific propensity score approach induces the same adjustment set, with comparable bias ($\approx 0.3$). Interacting $Z$ with time prior to applying de Chaisemartin and D'Haultfœuille's estimator is equivalent to adjusting for $Z_1$ alone, which yields a larger bias ($\approx 1.1$).

In Scenario \ref{ss:time-varying-effect}, we focus on $W_0$, a time-invariant covariate with a time-invariant effect on the outcome. Such a variable is typically not regarded as a confounder in \ac{DiD}, but as shown above it can constitute a sufficient adjustment set, and it does so here. We confirm that all estimators that effectively adjust for $W_0$ recover an unbiased estimate of the treatment effect (absolute bias $\approx 0.1$).

In Scenario \ref{ss:post-trt-vars3}, identification requires both the pre- and post-treatment covariate values, $\{W_0,W_1\}$. Most estimators cannot produce unbiased estimates without modification of the default adjustment set. The \ac{TWFE} model, de Chaisemartin and D'Haultfœuille's estimator and Stuart et al.~time-specific propensity score procedure each suffer from the issues just discussed, and the four-group variant has an unclear effective adjustment set. Heckman's outcome regression, Abadie's \ac{IPW} and Sant'Anna and Zhao's doubly robust estimators, all implemented within the Callaway and Sant'Anna package, exclude the post-treatment covariate value by default, yielding a bias of approximately $8.4$. By including the two time-constant copies carrying the pre- and post-treatment values, the correct adjustment is achieved and most estimators recover unbiased estimates (absolute bias $\approx 0$), although \ac{IPW}-based approaches exhibit somewhat larger bias, reaching up to $\approx 0.3$. As expected, including the pre-treatment value and covariate change yields comparable results.

Finally, in Scenario \ref{ss:mult-time-periods} (2), $W_1$ acts as a mediator at the first time period and as a confounder at the second. Therefore, including $W_1$ in the adjustment set biases the effect estimate at the first time period (absolute bias $\approx 4.5$), whereas the same variable is required to recover a valid estimate at the second time period, substantially reducing bias to $\approx 0.3$.

In summary, across all scenarios, whenever the effective adjustment set matches a sufficient adjustment set, bias is substantially eliminated across all estimation procedures.

\clearpage

\section{Discussion}\label{SecDis}

In this paper, we propose a graphical framework for variable selection in \ac{DiD} studies. The framework builds on the equivalence between the parallel trends and equi-confounding, a characterization that lends itself naturally to representation in a causal diagram. We show how this diagram can be simplified so that equi-confounding of unmeasured variables is represented by the absence of an edge, enabling direct application of standard graphical criteria like the backdoor criterion to select a minimally sufficient adjustment set that justifies conditional parallel trends.

Through a series of illustrative scenarios, we derived several insights that challenge conventional practice. We showed that unconditional parallel trends is a strong assumption requiring all confounding to cancel exactly, and that it can conflict with conditional parallel trends. This finding is particularly relevant in practice, where researchers often justify parallel trends on the basis of a pre-trends test using an unconditional model and then adjust for covariates as if doing so could only improve the estimate. Our analysis shows that if confounding is indeed perfectly offset unconditionally, conditioning on certain covariates can disrupt this balance and paradoxically \emph{introduce} bias.

Moreover, we demonstrated that a time-invariant covariate with a time-invariant effect on the outcome over time, which would not traditionally be considered a confounder in a \ac{DiD} study, can nonetheless be an important conditioning variable. This widens the pool of candidate adjustment variables, offering additional options when other covariates in sufficient adjustment sets are unavailable or poorly measured. We also clarified that post-treatment covariates, often excluded from adjustment by default, can play an essential role in identification in \ac{DiD} studies. The decision to adjust for post-treatment covariate values depends on the causal structure and, specifically, on what drives covariate changes over time.

When extending to multiple time periods, we distinguished between treatment type (static or dynamic) and rollout strategy (simultaneous or staggered), a distinction that has important implications for identification. For example, treatment-confounder feedback, which complicates covariate adjustment, arises only under dynamic treatment regimes and not merely because implementation is staggered.

Finally, we showed that several popular \ac{DiD} estimators handle covariates in ways such that the variables actually adjusted for can differ from those intended by the researcher. This divergence has not been explicitly recognized in the literature, where discussions of covariate handling in \ac{DiD} have instead focused on the choice of estimator. Our work suggests that this issue lies not in the estimator itself, but in a gap between what the estimator effectively conditions on and what identification requires. This gap arises in a range of widely used procedures. Closing the gap does not require switching estimators. Instead, modifying the covariate specification is sufficient to bring each estimator's effective adjustment set in line with the requirements of the causal model, recovering valid estimates without altering the estimation mechanics. Our simulation results reinforced this point by showing that sufficient adjustment sets consistently yielded low bias regardless of the estimation procedure, while insufficient sets produced substantial bias across all estimators. These results highlight that the choice of adjustment set deserves at least as much attention as the choice of estimator.

\newpage
\bibliographystyle{vancouver}
\bibliography{Bibliography}

\newpage
\appendix
\section{Identification via Wright's rules and Cramér's covariance formula in the context of linear SCMs}

\vspace{0.5cm}

\centering
\textbf{Scenario 5.2} (Figure \ref{cdscenario2})\\~\\

\justifying
\noindent Pairwise covariances via Wright's rules:\quad $\sigma(\Delta Y, A_1) = a + gijk - gih$
\vspace{-12pt}
\begin{flalign*}
& \hphantom{\text{Pairwise covariances via Wright's rules:}\quad} \sigma(\Delta Y, W_0) = ga - ih + ijk &&\\
& \hphantom{\text{Pairwise covariances via Wright's rules:}\quad} \sigma(\Delta Y, Z_0) = iga - h + jk &&\\
& \hphantom{\text{Pairwise covariances via Wright's rules:}\quad} \sigma(A_1, W_0) = g &&\\
& \hphantom{\text{Pairwise covariances via Wright's rules:}\quad} \sigma(A_1, Z_0) = gi &&
\end{flalign*}

\noindent Substituting into Cramér's covariance formula for the partial regression coefficient:
\begin{equation*}
\begin{split}
\beta (\Delta Y,A_1|W_0) & = \frac{\sigma(\Delta Y, A_1) - \sigma(\Delta Y,W_0) \sigma(A_1,W_0)}{1-\sigma(A_1,W_0)^2} \\
& = \frac{a+gijk-gih - (ga-ih+ijk)(g)}{1-g^2} \\
& = \frac{a+gijk-gih - ag^2 +gih -gijk}{1-g^2} \\
& = \frac{a - ag^2}{1-g^2} \\
& = \frac{a(1-g^2)}{1-g^2} \\
& = a\;
\end{split}
\end{equation*}

\begin{equation*}
\begin{split}
\beta (\Delta Y,A_1|Z_0) & = \frac{\sigma(\Delta Y, A_1) - \sigma(\Delta Y,Z_0) \sigma(A_1,Z_0)}{1-\sigma(A_1,Z_0)^2} \\
& = \frac{a+gijk-gih - (iga-h+jk) (gi)}{1-(gi)^2} \\
& = \frac{a+gijk-gih - a(gi)^2 +gih -gijk}{1-(gi)^2} \\
& = \frac{a - a(gi)^2}{1-(gi)^2} \\
& = \frac{a(1-(gi)^2)}{1-(gi)^2} \\
& = a\
\end{split}
\end{equation*}

\newpage

\centering
\textbf{Scenario 5.3.1} (Figure \ref{cdscenario3a})\\~\\

\justifying
\noindent Pairwise covariances via Wright's rules:\quad $\sigma(\Delta Y, A_1) = a + gih - gf$
\vspace{-12pt}
\begin{flalign*}
& \hphantom{\text{Pairwise covariances via Wright's rules:}\quad} \sigma(\Delta Y, W_0) = ga + ih - f &&\\
& \hphantom{\text{Pairwise covariances via Wright's rules:}\quad} \sigma(A_1, W_0) = g &&
\end{flalign*}

\noindent Substituting into Cramér's covariance formula for the partial regression coefficient:
\begin{equation*}
\begin{split}
\beta (\Delta Y,A_1|W_0) & = \frac{\sigma(\Delta Y, A_1) - \sigma(\Delta Y,W_0) \sigma(A_1,W_0)}{1-\sigma(A_1,W_0)^2} \\
& = \frac{a+gih-gf - (ga+ih-f) (g)}{1-g^2} \\
& = \frac{a+gih-gf - ag^2 -gih +gf}{1-g^2} \\
& = \frac{a-ag^2}{1-g^2} \\
& = \frac{a(1-g^2)}{1-g^2} \\
& = a\;
\end{split}
\end{equation*}

\newpage
\centering
\textbf{Scenario 5.3.2} (Figure \ref{cdscenario3b})\\~\\

\justifying
\noindent Pairwise covariances via Wright's rules:\quad $\sigma(\Delta Y, A_1) = a - gf + gkh + ijh$
\vspace{-12pt}
\begin{flalign*}
& \hphantom{\text{Pairwise covariances via Wright's rules:}\quad} \sigma(\Delta Y, W_0) = -f + ga + kh &&\\
& \hphantom{\text{Pairwise covariances via Wright's rules:}\quad} \sigma(\Delta Y, W_1) = h - kf + kga + jia &&\\
& \hphantom{\text{Pairwise covariances via Wright's rules:}\quad} \sigma(\Delta Y, Z_0) = ia + jh &&\\
& \hphantom{\text{Pairwise covariances via Wright's rules:}\quad} \sigma(A_1, W_0) = g &&\\
& \hphantom{\text{Pairwise covariances via Wright's rules:}\quad} \sigma(A_1, W_1) = gk + ij &&\\
& \hphantom{\text{Pairwise covariances via Wright's rules:}\quad} \sigma(A_1, Z_0) = i &&\\
& \hphantom{\text{Pairwise covariances via Wright's rules:}\quad} \sigma(W_0, W_1) = k &&\\
& \hphantom{\text{Pairwise covariances via Wright's rules:}\quad} \sigma(W_0, Z_0) = 0 &&
\end{flalign*}

\noindent Conditional standard deviations given $W_0$:\quad $\sigma(\Delta Y | W_0) = \left[1 - \sigma(\Delta Y, W_0)^2\right]^{1/2} = \left[1 - (-f + ga + kh)^2\right]^{1/2}$
\vspace{-10pt}
\begin{flalign*}
& \hphantom{\text{Conditional standard deviations given $W_0$:}\quad} \sigma(A_1 | W_0) = \left[1 - \sigma(A_1, W_0)^2\right]^{1/2} = \left[1 - g^2\right]^{1/2} &&
\end{flalign*}

\noindent Partial correlations given $W_0$:\quad $\displaystyle\rho(\Delta Y, A_1 | W_0) = \frac{\sigma(\Delta Y, A_1) - \sigma(\Delta Y, W_0)\sigma(A_1, W_0)}{\left[1 - \sigma(\Delta Y, W_0)^2\right]^{1/2}\left[1 - \sigma(A_1, W_0)^2\right]^{1/2}}$
\vspace{-5pt}
\begin{flalign*}
& \hphantom{\text{Partial correlations given $W_0$:}\quad\rho(\Delta Y, A_1 | W_0)} = \frac{(a - gf + gkh + ijh) - (-f + ga + kh)(g)}{\left[1 - (-f + ga + kh)^2\right]^{1/2}\left[1 - g^2\right]^{1/2}} &&\\[6pt]
& \hphantom{\text{Partial correlations given $W_0$:}\quad\rho(\Delta Y, A_1 | W_0)} = \frac{a(1 - g^2) + ijh}{\left[1 - (-f + ga + kh)^2\right]^{1/2}\left[1 - g^2\right]^{1/2}} &&
\end{flalign*}
\vspace{-12pt}
\begin{flalign*}
& \hphantom{\text{Partial correlations given $W_0$:}\quad} \rho(\Delta Y, W_1 | W_0) = \frac{\sigma(\Delta Y, W_1) - \sigma(\Delta Y, W_0)\sigma(W_1, W_0)}{\left[1 - \sigma(\Delta Y, W_0)^2\right]^{1/2}\left[1 - \sigma(W_1, W_0)^2\right]^{1/2}} &&\\[6pt]
& \hphantom{\text{Partial correlations given $W_0$:}\quad\rho(\Delta Y, W_1 | W_0)} = \frac{(h - kf + kga + jia) - (-f + ga + kh)(k)}{\left[1 - (-f + ga + kh)^2\right]^{1/2}\left[1 - k^2\right]^{1/2}} &&\\[6pt]
& \hphantom{\text{Partial correlations given $W_0$:}\quad\rho(\Delta Y, W_1 | W_0)} = \frac{h(1 - k^2) + jia}{\left[1 - (-f + ga + kh)^2\right]^{1/2}\left[1 - k^2\right]^{1/2}} &&
\end{flalign*}
\vspace{-12pt}
\begin{flalign*}
& \hphantom{\text{Partial correlations given $W_0$:}\quad} \rho(A_1, W_1 | W_0) = \frac{\sigma(A_1, W_1) - \sigma(A_1, W_0)\sigma(W_1, W_0)}{\left[1 - \sigma(A_1, W_0)^2\right]^{1/2}\left[1 - \sigma(W_1, W_0)^2\right]^{1/2}} &&\\[6pt]
& \hphantom{\text{Partial correlations given $W_0$:}\quad\rho(A_1, W_1 | W_0)} = \frac{(gk + ij) - (g)(k)}{\left[1 - g^2\right]^{1/2}\left[1 - k^2\right]^{1/2}} &&\\[6pt]
& \hphantom{\text{Partial correlations given $W_0$:}\quad\rho(A_1, W_1 | W_0)} = \frac{ij}{\left[1 - g^2\right]^{1/2}\left[1 - k^2\right]^{1/2}} &&
\end{flalign*}
\vspace{-12pt}
\begin{flalign*}
& \hphantom{\text{Partial correlations given $W_0$:}\quad} \rho(\Delta Y, Z_0 | W_0) = \frac{\sigma(\Delta Y, Z_0) - \sigma(\Delta Y, W_0)\sigma(Z_0, W_0)}{\left[1 - \sigma(\Delta Y, W_0)^2\right]^{1/2}\left[1 - \sigma(Z_0, W_0)^2\right]^{1/2}} &&\\[6pt]
& \hphantom{\text{Partial correlations given $W_0$:}\quad\rho(\Delta Y, Z_0 | W_0)} = \frac{(ia + jh) - (-f + ga + kh)(0)}{\left[1 - (-f + ga + kh)^2\right]^{1/2}\left[1 - 0\right]^{1/2}} &&\\[6pt]
& \hphantom{\text{Partial correlations given $W_0$:}\quad\rho(\Delta Y, Z_0 | W_0)} = \frac{ia + jh}{\left[1 - (-f + ga + kh)^2\right]^{1/2}} &&
\end{flalign*}
\vspace{-12pt}
\begin{flalign*}
& \hphantom{\text{Partial correlations given $W_0$:}\quad} \rho(A_1, Z_0 | W_0) = \frac{\sigma(A_1, Z_0) - \sigma(A_1, W_0)\sigma(Z_0, W_0)}{\left[1 - \sigma(A_1, W_0)^2\right]^{1/2}\left[1 - \sigma(Z_0, W_0)^2\right]^{1/2}} &&\\[6pt]
& \hphantom{\text{Partial correlations given $W_0$:}\quad\rho(A_1, Z_0 | W_0)} = \frac{(i) - (g)(0)}{\left[1 - g^2\right]^{1/2}\left[1 - 0\right]^{1/2}} &&\\[6pt]
& \hphantom{\text{Partial correlations given $W_0$:}\quad\rho(A_1, Z_0 | W_0)} = \frac{i}{\left[1 - g^2\right]^{1/2}} &&
\end{flalign*}

\vspace{0.5cm}

\noindent Substituting into Cramér's covariance formula for the partial regression coefficient:
{\small
\begin{equation*}
\begin{split}
\beta(\Delta Y, A_1 | Z_0, W_0) & = \frac{\sigma(\Delta Y | W_0)}{\sigma(A_1 | W_0)} \cdot \frac{\rho(\Delta Y, A_1 | W_0) - \rho(\Delta Y, Z_0 | W_0)\,\rho(A_1, Z_0 | W_0)}{1 - \rho(A_1, Z_0 | W_0)^2} \\[10pt]
& = \frac{\left[1 - (-f+ga+kh)^2\right]^{1/2}}{\left[1-g^2\right]^{1/2}} \cdot \frac{\dfrac{a(1-g^2)+ijh}{\left[1-(-f+ga+kh)^2\right]^{1/2}\left[1-g^2\right]^{1/2}} - \dfrac{ia+jh}{\left[1-(-f+ga+kh)^2\right]^{1/2}} \cdot \dfrac{i}{\left[1-g^2\right]^{1/2}}}{1 - \dfrac{i^2}{1-g^2}} \\[10pt]
& = \frac{\left[1 - (-f+ga+kh)^2\right]^{1/2}}{\left[1-g^2\right]^{1/2}} \cdot \frac{\dfrac{a(1-g^2)+ijh - i^2a - ijh}{\left[1-(-f+ga+kh)^2\right]^{1/2}\left[1-g^2\right]^{1/2}}}{\dfrac{1-g^2-i^2}{1-g^2}} \\[10pt]
& = \frac{\left[1 - (-f+ga+kh)^2\right]^{1/2}}{\left[1-g^2\right]^{1/2}} \cdot \frac{a(1-g^2) - i^2a}{\left[1-(-f+ga+kh)^2\right]^{1/2}\left[1-g^2\right]^{1/2}} \cdot \frac{1-g^2}{1-g^2-i^2} \\[10pt]
& = \frac{1}{1-g^2} \cdot \frac{a(1-g^2-i^2)}{1} \cdot \frac{1-g^2}{1-g^2-i^2} \\[10pt]
& = \frac{1}{(1-g^2)} \cdot a(1-g^2) \\[10pt]
& = a
\end{split}
\end{equation*}
}

\begin{landscape}
\begin{equation*}
\begin{split}
\beta(\Delta Y, A_1 | W_1, W_0) & = \frac{\sigma(\Delta Y | W_0)}{\sigma(A_1 | W_0)} \cdot \frac{\rho(\Delta Y, A_1 | W_0) - \rho(\Delta Y, W_1 | W_0)\,\rho(A_1, W_1 | W_0)}{1 - \rho(A_1, W_1 | W_0)^2} \\[10pt]
& = \frac{\left[1 - (-f+ga+kh)^2\right]^{1/2}}{\left[1-g^2\right]^{1/2}} \cdot \frac{\dfrac{a(1-g^2)+ijh}{\left[1-(-f+ga+kh)^2\right]^{1/2}\left[1-g^2\right]^{1/2}} - \dfrac{h(1-k^2)+jia}{\left[1-(-f+ga+kh)^2\right]^{1/2}\left[1-k^2\right]^{1/2}} \cdot \dfrac{ij}{\left[1-g^2\right]^{1/2}\left[1-k^2\right]^{1/2}}}{1 - \dfrac{i^2j^2}{(1-g^2)(1-k^2)}} \\[10pt]
& = \frac{\left[1 - (-f+ga+kh)^2\right]^{1/2}}{\left[1-g^2\right]^{1/2}} \cdot \frac{\dfrac{\left[a(1-g^2)+ijh\right](1-k^2) - \left[h(1-k^2)+jia\right] \cdot ij}{\left[1-(-f+ga+kh)^2\right]^{1/2}\left[1-g^2\right]^{1/2}(1-k^2)}}{\dfrac{(1-g^2)(1-k^2)-i^2j^2}{(1-g^2)(1-k^2)}} \\[10pt]
& = \frac{\left[1 - (-f+ga+kh)^2\right]^{1/2}}{\left[1-g^2\right]^{1/2}} \cdot \frac{\dfrac{a(1-g^2)(1-k^2)+ijh(1-k^2) - ijh(1-k^2) - ai^2j^2}{\left[1-(-f+ga+kh)^2\right]^{1/2}\left[1-g^2\right]^{1/2}(1-k^2)}}{\dfrac{(1-g^2)(1-k^2)-i^2j^2}{(1-g^2)(1-k^2)}} \\[10pt]
& = \frac{\left[1 - (-f+ga+kh)^2\right]^{1/2}}{\left[1-g^2\right]^{1/2}} \cdot \frac{\dfrac{a\left[(1-g^2)(1-k^2) - i^2j^2\right]}{\left[1-(-f+ga+kh)^2\right]^{1/2}\left[1-g^2\right]^{1/2}(1-k^2)}}{\dfrac{(1-g^2)(1-k^2)-i^2j^2}{(1-g^2)(1-k^2)}} \\[10pt]
& = \frac{\left[1 - (-f+ga+kh)^2\right]^{1/2}}{\left[1-g^2\right]^{1/2}} \cdot \frac{a\left[(1-g^2)(1-k^2) - i^2j^2\right]}{\left[1-(-f+ga+kh)^2\right]^{1/2}\left[1-g^2\right]^{1/2}(1-k^2)} \cdot \frac{(1-g^2)(1-k^2)}{(1-g^2)(1-k^2)-i^2j^2} \\[10pt]
& = \frac{1}{1-g^2} \cdot a(1-g^2) \\[10pt]
& = a
\end{split}
\end{equation*}
\end{landscape}

\newpage
\centering
\textbf{Scenario 5.3.3} (Figure \ref{cdscenario3c})\\~\\
\justifying
\noindent Pairwise covariances via Wright's rules:\quad $\sigma(\Delta Y, A_1) = a - gf + gjh + bih$
\vspace{-12pt}
\begin{flalign*}
& \hphantom{\text{Pairwise covariances via Wright's rules:}\quad} \sigma(\Delta Y, W_0) = -f + ga + jh &&\\
& \hphantom{\text{Pairwise covariances via Wright's rules:}\quad} \sigma(\Delta Y, W_1) = h - jf + jga + iba &&\\
& \hphantom{\text{Pairwise covariances via Wright's rules:}\quad} \sigma(A_1, W_0) = g &&\\
& \hphantom{\text{Pairwise covariances via Wright's rules:}\quad} \sigma(A_1, W_1) = gj + bi &&\\
& \hphantom{\text{Pairwise covariances via Wright's rules:}\quad} \sigma(W_0, W_1) = j &&
\end{flalign*}

\noindent Conditional standard deviations given $W_0$:\quad $\sigma(\Delta Y | W_0) = \left[1 - \sigma(\Delta Y, W_0)^2\right]^{1/2} = \left[1 - (-f + ga + jh)^2\right]^{1/2}$
\vspace{-10pt}
\begin{flalign*}
& \hphantom{\text{Conditional standard deviations given $W_0$:}\quad} \sigma(A_1 | W_0) = \left[1 - \sigma(A_1, W_0)^2\right]^{1/2} = \left[1 - g^2\right]^{1/2} &&
\end{flalign*}

\noindent Partial correlations given $W_0$:\quad $\displaystyle\rho(\Delta Y, A_1 | W_0) = \frac{\sigma(\Delta Y, A_1) - \sigma(\Delta Y, W_0)\sigma(A_1, W_0)}{\left[1 - \sigma(\Delta Y, W_0)^2\right]^{1/2}\left[1 - \sigma(A_1, W_0)^2\right]^{1/2}}$
\vspace{-5pt}
\begin{flalign*}
& \hphantom{\text{Partial correlations given $W_0$:}\quad\rho(\Delta Y, A_1 | W_0)} = \frac{(a - gf + gjh + bih) - (-f + ga + jh)(g)}{\left[1 - (-f + ga + jh)^2\right]^{1/2}\left[1 - g^2\right]^{1/2}} &&\\[6pt]
& \hphantom{\text{Partial correlations given $W_0$:}\quad\rho(\Delta Y, A_1 | W_0)} = \frac{a(1 - g^2) + bih}{\left[1 - (-f + ga + jh)^2\right]^{1/2}\left[1 - g^2\right]^{1/2}} &&
\end{flalign*}
\vspace{-12pt}
\begin{flalign*}
& \hphantom{\text{Partial correlations given $W_0$:}\quad} \rho(\Delta Y, W_1 | W_0) = \frac{\sigma(\Delta Y, W_1) - \sigma(\Delta Y, W_0)\sigma(W_1, W_0)}{\left[1 - \sigma(\Delta Y, W_0)^2\right]^{1/2}\left[1 - \sigma(W_1, W_0)^2\right]^{1/2}} &&\\[6pt]
& \hphantom{\text{Partial correlations given $W_0$:}\quad\rho(\Delta Y, W_1 | W_0)} = \frac{(h - jf + jga + iba) - (-f + ga + jh)(j)}{\left[1 - (-f + ga + jh)^2\right]^{1/2}\left[1 - j^2\right]^{1/2}} &&\\[6pt]
& \hphantom{\text{Partial correlations given $W_0$:}\quad\rho(\Delta Y, W_1 | W_0)} = \frac{h(1 - j^2) + iba}{\left[1 - (-f + ga + jh)^2\right]^{1/2}\left[1 - j^2\right]^{1/2}} &&
\end{flalign*}
\vspace{-12pt}
\begin{flalign*}
& \hphantom{\text{Partial correlations given $W_0$:}\quad} \rho(A_1, W_1 | W_0) = \frac{\sigma(A_1, W_1) - \sigma(A_1, W_0)\sigma(W_1, W_0)}{\left[1 - \sigma(A_1, W_0)^2\right]^{1/2}\left[1 - \sigma(W_1, W_0)^2\right]^{1/2}} &&\\[6pt]
& \hphantom{\text{Partial correlations given $W_0$:}\quad\rho(A_1, W_1 | W_0)} = \frac{(gj + bi) - (g)(j)}{\left[1 - g^2\right]^{1/2}\left[1 - j^2\right]^{1/2}} &&\\[6pt]
& \hphantom{\text{Partial correlations given $W_0$:}\quad\rho(A_1, W_1 | W_0)} = \frac{bi}{\left[1 - g^2\right]^{1/2}\left[1 - j^2\right]^{1/2}} &&
\end{flalign*}

\vspace{0.5cm}

\begin{landscape}
\noindent Substituting into Cramér's covariance formula for the partial regression coefficient:
\begin{equation*}
\begin{split}
\beta(\Delta Y, A_1 | W_1, W_0) & = \frac{\sigma(\Delta Y | W_0)}{\sigma(A_1 | W_0)} \cdot \frac{\rho(\Delta Y, A_1 | W_0) - \rho(\Delta Y, W_1 | W_0)\,\rho(A_1, W_1 | W_0)}{1 - \rho(A_1, W_1 | W_0)^2} \\[10pt]
& = \frac{\left[1 - (-f+ga+jh)^2\right]^{1/2}}{\left[1-g^2\right]^{1/2}} \cdot \frac{\dfrac{a(1-g^2)+bih}{\left[1-(-f+ga+jh)^2\right]^{1/2}\left[1-g^2\right]^{1/2}} - \dfrac{h(1-j^2)+iba}{\left[1-(-f+ga+jh)^2\right]^{1/2}\left[1-j^2\right]^{1/2}} \cdot \dfrac{bi}{\left[1-g^2\right]^{1/2}\left[1-j^2\right]^{1/2}}}{1 - \dfrac{b^2i^2}{(1-g^2)(1-j^2)}} \\[10pt]
& = \frac{\left[1 - (-f+ga+jh)^2\right]^{1/2}}{\left[1-g^2\right]^{1/2}} \cdot \frac{\dfrac{\left[a(1-g^2)+bih\right](1-j^2) - \left[h(1-j^2)+iba\right] \cdot bi}{\left[1-(-f+ga+jh)^2\right]^{1/2}\left[1-g^2\right]^{1/2}(1-j^2)}}{\dfrac{(1-g^2)(1-j^2)-b^2i^2}{(1-g^2)(1-j^2)}} \\[10pt]
& = \frac{\left[1 - (-f+ga+jh)^2\right]^{1/2}}{\left[1-g^2\right]^{1/2}} \cdot \frac{\dfrac{a(1-g^2)(1-j^2)+bih(1-j^2) - bih(1-j^2) - ab^2i^2}{\left[1-(-f+ga+jh)^2\right]^{1/2}\left[1-g^2\right]^{1/2}(1-j^2)}}{\dfrac{(1-g^2)(1-j^2)-b^2i^2}{(1-g^2)(1-j^2)}} \\[10pt]
& = \frac{\left[1 - (-f+ga+jh)^2\right]^{1/2}}{\left[1-g^2\right]^{1/2}} \cdot \frac{\dfrac{a\left[(1-g^2)(1-j^2) - b^2i^2\right]}{\left[1-(-f+ga+jh)^2\right]^{1/2}\left[1-g^2\right]^{1/2}(1-j^2)}}{\dfrac{(1-g^2)(1-j^2)-b^2i^2}{(1-g^2)(1-j^2)}} \\[10pt]
& = \frac{\left[1 - (-f+ga+jh)^2\right]^{1/2}}{\left[1-g^2\right]^{1/2}} \cdot \frac{a\left[(1-g^2)(1-j^2) - b^2i^2\right]}{\left[1-(-f+ga+jh)^2\right]^{1/2}\left[1-g^2\right]^{1/2}(1-j^2)} \cdot \frac{(1-g^2)(1-j^2)}{(1-g^2)(1-j^2)-b^2i^2} \\[10pt]
& = \frac{1}{1-g^2} \cdot a(1-g^2) \\[10pt]
& = a
\end{split}
\end{equation*}
\end{landscape}

\newpage
\centering
\textbf{Scenario 5.3.4} (Figure \ref{cdscenario3d})\\~\\

\justifying
\noindent Pairwise covariances via Wright's rules:\quad $\sigma(\Delta Y, A_1) = a + ih + gjh - gf$
\vspace{-12pt}
\begin{flalign*}
& \hphantom{\text{Pairwise covariances via Wright's rules:}\quad} \sigma(\Delta Y, W_0) = ga + gih + jh - f &&\\
& \hphantom{\text{Pairwise covariances via Wright's rules:}\quad} \sigma(A_1, W_0) = g &&
\end{flalign*}

\noindent Substituting into Cramér's covariance formula for the partial regression coefficient:
\begin{equation*}
\begin{split}
\beta (\Delta Y,A_1|W_0) & = \frac{\sigma(\Delta Y, A_1) - \sigma(\Delta Y,W_0) \sigma(A_1,W_0)}{1-\sigma(A_1,W_0)^2} \\
& = \frac{a+ih+gjh-gf - (ga+gih+jh-f) (g)}{1-g^2} \\
& = \frac{a+ih+gjh-gf - ag^2 -ihg^2 -gjh +gf}{1-g^2} \\
& = \frac{a-ag^2+ih-ihg^2}{1-g^2} \\
& = \frac{a-ag^2}{1-g^2}+\frac{ih-ihg^2}{1-g^2} \\
& = \frac{a(1-g^2)}{1-g^2}+\frac{ih(1-g^2)}{1-g^2} \\
& = a+ih \text{ (total effect)}\;
\end{split}
\end{equation*}

\newpage
\centering
\textbf{Scenario 5.4} (Figure \ref{cdscenario4})\\~\\

\justifying
\noindent Pairwise covariances via Wright's rules:\quad $\sigma(\Delta Y_1, A_1) = a + gih - gf$
\vspace{-12pt}
\begin{flalign*}
& \hphantom{\text{Pairwise covariances via Wright's rules:}\quad} \sigma(\Delta Y_2, A_1) = a + gijk - gf &&\\
& \hphantom{\text{Pairwise covariances via Wright's rules:}\quad} \sigma(\Delta Y_1, W_0) = ga + ih - f &&\\
& \hphantom{\text{Pairwise covariances via Wright's rules:}\quad} \sigma(\Delta Y_2, W_0) = ga + ijk - f &&\\
& \hphantom{\text{Pairwise covariances via Wright's rules:}\quad} \sigma(A_1, W_0) = g &&
\end{flalign*}

\noindent Substituting into Cramér's covariance formula for the partial regression coefficient at times 1 and 2:
\begin{equation*}
\begin{split}
\beta (\Delta Y_1,A_1|W_0) & = \frac{\sigma(\Delta Y_1, A_1) - \sigma(\Delta Y_1,W_0) \sigma(A_1,W_0)}{1-\sigma(A_1,W_0)^2} \\
& = \frac{a+gih-gf - (ga+ih-f) (g)}{1-g^2} \\
& = \frac{a+gih-gf - ag^2 -gih +gf}{1-g^2} \\
& = \frac{a-ag^2}{1-g^2} \\
& = \frac{a(1-g^2)}{1-g^2} \\
& = a\;
\end{split}
\end{equation*}

\begin{equation*}
\begin{split}
\beta (\Delta Y_2,A_1|W_0) & = \frac{\sigma(\Delta Y_2, A_1) - \sigma(\Delta Y_2,W_0) \sigma(A_1,W_0)}{1-\sigma(A_1,W_0)^2} \\
& = \frac{a+gijk-gf - (ga+ijk-f) (g)}{1-g^2} \\
& = \frac{a+gijk-gf - ag^2 -gijk +gf}{1-g^2} \\
& = \frac{a-ag^2}{1-g^2} \\
& = \frac{a(1-g^2)}{1-g^2} \\
& = a\;
\end{split}
\end{equation*}

\newpage
\begin{landscape}  

\centering
\textbf{Scenario 5.4 (2)} (Figure \ref{cdscenario4_2})\\~\\
\justifying

\noindent Pairwise covariances via Wright's rules:\quad $\sigma(\Delta Y_1, A_1) = a + hi + gji -gf$
\vspace{-12pt}
\begin{flalign*}
& \hphantom{\text{Pairwise covariances via Wright's rules:}\quad} \sigma(\Delta Y_1,W_0) = ga + ji + ghi -f &&\\
& \hphantom{\text{Pairwise covariances via Wright's rules:}\quad} \sigma(A_1,W_0) = g &&\\
& \hphantom{\text{Pairwise covariances via Wright's rules:}\quad} \sigma(\Delta Y_2, A_2) = a + mn - fgp - fjl - fghl + nkl + nkhp + nkgjp + b^2hkn &&\\
& \hphantom{\text{Pairwise covariances via Wright's rules:}\quad} \sigma(\Delta Y_2, W_0) = -f + agp + ajl + aghl + mngp + mnjl + mnghl + nkj + nkgh &&\\
& \hphantom{\text{Pairwise covariances via Wright's rules:}\quad} \sigma(\Delta Y_2, W_1) = nk + al + ahp + ab^2h + agjp + mnl + mnhp + mnb^2h + mngjp - fj - fgh &&\\
& \hphantom{\text{Pairwise covariances via Wright's rules:}\quad} \sigma(A_2, W_0) = gp + jl + ghl &&\\
& \hphantom{\text{Pairwise covariances via Wright's rules:}\quad} \sigma(A_2, W_1) = l + hp + b^2h + gjp &&\\
& \hphantom{\text{Pairwise covariances via Wright's rules:}\quad} \sigma(W_0, W_1) = j + gh &&
\end{flalign*}

\noindent Conditional standard deviations given $W_0$:\quad $\sigma(\Delta Y_2 | W_0) = \left[1 - (-f + agp + ajl + aghl + mngp + mnjl + mnghl + nkj + nkgh)^2\right]^{1/2}$
\vspace{-10pt}
\begin{flalign*}
& \hphantom{\text{Conditional standard deviations given $W_0$:}\quad} \sigma(A_2 | W_0) = \left[1 - (gp + jl + ghl)^2\right]^{1/2} &&
\end{flalign*}

\noindent Partial correlations given $W_0$:
\begin{equation*}
\begin{split}
\rho(\Delta Y_2, A_2 | W_0) & = \frac{\sigma(\Delta Y_2, A_2) - \sigma(\Delta Y_2, W_0)\sigma(A_2, W_0)}{\left[1 - \sigma(\Delta Y_2, W_0)^2\right]^{1/2}\left[1 - \sigma(A_2, W_0)^2\right]^{1/2}} \\[6pt]
& = \frac{(a + mn - fgp - fjl - fghl + nkl + nkhp + nkgjp + b^2hkn) - (-f + agp + ajl + aghl + mngp + mnjl + mnghl + nkj + nkgh)(gp + jl + ghl)}{\left[1 - (-f + agp + ajl + aghl + mngp + mnjl + mnghl + nkj + nkgh)^2\right]^{1/2}\left[1 - (gp + jl + ghl)^2\right]^{1/2}} \\[6pt]
& = \frac{(a + mn)\left[1 - (gp + jl + ghl)^2\right] + nk\left[(l + hp + b^2h + gjp) - (gp + jl + ghl)(j + gh)\right]}{\left[1 - (-f + agp + ajl + aghl + mngp + mnjl + mnghl + nkj + nkgh)^2\right]^{1/2}\left[1 - (gp + jl + ghl)^2\right]^{1/2}}
\end{split}
\end{equation*}

\begin{equation*}
\begin{split}
\rho(\Delta Y_2, W_1 | W_0) & = \frac{\sigma(\Delta Y_2, W_1) - \sigma(\Delta Y_2, W_0)\sigma(W_1, W_0)}{\left[1 - \sigma(\Delta Y_2, W_0)^2\right]^{1/2}\left[1 - \sigma(W_1, W_0)^2\right]^{1/2}} \\[6pt]
& = \scalebox{0.93}{$\displaystyle\frac{(nk + al + ahp + ab^2h + agjp + mnl + mnhp + mnb^2h + mngjp - fj - fgh) - (-f + agp + ajl + aghl + mngp + mnjl + mnghl + nkj + nkgh)(j + gh)}{\left[1 - (-f + agp + ajl + aghl + mngp + mnjl + mnghl + nkj + nkgh)^2\right]^{1/2}\left[1 - (j + gh)^2\right]^{1/2}}$} \\[6pt]
& = \frac{nk\left[1 - (j + gh)^2\right] + (a + mn)\left[(l + hp + b^2h + gjp) - (gp + jl + ghl)(j + gh)\right]}{\left[1 - (-f + agp + ajl + aghl + mngp + mnjl + mnghl + nkj + nkgh)^2\right]^{1/2}\left[1 - (j + gh)^2\right]^{1/2}} \\[30pt]
\rho(A_2, W_1 | W_0) & = \frac{\sigma(A_2, W_1) - \sigma(A_2, W_0)\sigma(W_1, W_0)}{\left[1 - \sigma(A_2, W_0)^2\right]^{1/2}\left[1 - \sigma(W_1, W_0)^2\right]^{1/2}} = \frac{(l + hp + b^2h + gjp) - (gp + jl + ghl)(j + gh)}{\left[1 - (gp + jl + ghl)^2\right]^{1/2}\left[1 - (j + gh)^2\right]^{1/2}}
\end{split}
\end{equation*}

\vspace{1cm}

\noindent Substituting into Cram\'er's covariance formula for the partial regression coefficient at time 1:
\begin{equation*}
\begin{split}
\beta (\Delta Y_1,A_1|W_0) & = \frac{\sigma(\Delta Y_1, A_1) - \sigma(\Delta Y_1,W_0) \sigma(A_1,W_0)}{1-\sigma(A_1,W_0)^2} \\
& = \frac{a + hi + gji -gf - (ga + ji + ghi -f)(g)}{1-g^2} \\
& = \frac{a + hi + gji -gf - ag^2 -gji -hig^2 +gf}{1-g^2} \\
& = \frac{a - ag^2 + hi -hig^2}{1-g^2} \\
& = \frac{a(1-g^2)}{1-g^2} + \frac{hi (1-g^2)}{1-g^2} \\
& = a + hi \text{ (total effect)}\;
\end{split}
\end{equation*}

\noindent Simplifying the main fraction of Cram\'er's covariance formula for the partial regression coefficient at time 2:
\begin{equation*}
\begin{split}
& \frac{\rho(\Delta Y_2, A_2 | W_0) - \rho(\Delta Y_2, W_1 | W_0)\,\rho(A_2, W_1 | W_0)}{1 - \rho(A_2, W_1 | W_0)^2} \\[10pt]
& = \Bigg(\frac{1}{\left[1-(-f + agp + ajl + aghl + mngp + mnjl + mnghl + nkj + nkgh)^2\right]^{1/2}} \cdot \\[6pt]
& \quad \frac{\left\{(a+mn)\left[1-(gp+jl+ghl)^2\right] + nk\left[(l+hp+b^2h+gjp) - (gp+jl+ghl)(j+gh)\right]\right\}\left[1-(j+gh)^2\right]}{\left[1-(gp+jl+ghl)^2\right]^{1/2}\left[1-(j+gh)^2\right]} \\[6pt]
& - \frac{1}{\left[1-(-f + agp + ajl + aghl + mngp + mnjl + mnghl + nkj + nkgh)^2\right]^{1/2}} \cdot \\[6pt]
& \quad \frac{\left\{nk\left[1-(j+gh)^2\right] + (a+mn)\left[(l+hp+b^2h+gjp) - (gp+jl+ghl)(j+gh)\right]\right\}\left[(l+hp+b^2h+gjp) - (gp+jl+ghl)(j+gh)\right]}{\left[1-(gp+jl+ghl)^2\right]^{1/2}\left[1-(j+gh)^2\right]}\Bigg) \\[6pt]
& \div \frac{\left[1-(gp+jl+ghl)^2\right]\left[1-(j+gh)^2\right] - \left[(l+hp+b^2h+gjp) - (gp+jl+ghl)(j+gh)\right]^2}{\left[1-(gp+jl+ghl)^2\right]\left[1-(j+gh)^2\right]} \\[10pt]
& = \frac{(a+mn)\left\{\left[1-(gp+jl+ghl)^2\right]\left[1-(j+gh)^2\right] - \left[(l+hp+b^2h+gjp) - (gp+jl+ghl)(j+gh)\right]^2\right\}}{\left[1-(-f + agp + ajl + aghl + mngp + mnjl + mnghl + nkj + nkgh)^2\right]^{1/2}\left[1-(gp+jl+ghl)^2\right]^{1/2}\left[1-(j+gh)^2\right]} \\[6pt]
& \quad \cdot \frac{\left[1-(gp+jl+ghl)^2\right]\left[1-(j+gh)^2\right]}{\left[1-(gp+jl+ghl)^2\right]\left[1-(j+gh)^2\right] - \left[(l+hp+b^2h+gjp) - (gp+jl+ghl)(j+gh)\right]^2} \\[10pt]
& = \frac{(a+mn)\left[1-(gp+jl+ghl)^2\right]}{\left[1-(-f + agp + ajl + aghl + mngp + mnjl + mnghl + nkj + nkgh)^2\right]^{1/2}\left[1-(gp+jl+ghl)^2\right]^{1/2}}
\end{split}
\end{equation*}

\vspace{0.5cm}

\noindent Substituting back into Cram\'er's covariance formula for the partial regression coefficient at time 2:
\begin{equation*}
\begin{split}
\beta(\Delta Y_2, A_2 | W_1, W_0) & = \frac{\sigma(\Delta Y_2 | W_0)}{\sigma(A_2 | W_0)} \cdot \frac{\rho(\Delta Y_2, A_2 | W_0) - \rho(\Delta Y_2, W_1 | W_0)\,\rho(A_2, W_1 | W_0)}{1 - \rho(A_2, W_1 | W_0)^2} \\
& = \frac{\left[1 - (-f + agp + ajl + aghl + mngp + mnjl + mnghl + nkj + nkgh)^2\right]^{1/2}}{\left[1-(gp+jl+ghl)^2\right]^{1/2}} \\
& \quad \cdot \frac{(a+mn)\left[1-(gp+jl+ghl)^2\right]}{\left[1-(-f + agp + ajl + aghl + mngp + mnjl + mnghl + nkj + nkgh)^2\right]^{1/2}\left[1-(gp+jl+ghl)^2\right]^{1/2}} \\
& = \frac{1}{1-(gp+jl+ghl)^2} \cdot (a+mn) \cdot \left[1-(gp+jl+ghl)^2\right] \\
& = a + mn \text{ (total effect)}\;
\end{split}
\end{equation*}
\end{landscape}

\newpage
\section{Estimation}

Below, we detail our implementation of all estimation methods in common notation, emphasizing how each incorporates covariates. As a shorthand for the adjustment set, we use $\mathbf{X}_{i}$ for the vector of covariates observed for unit $i$, which may include time-varying covariates. In our simulations, the components of this vector correspond to elements from the graphical models, such as $W_{i0}$, $Z_{i0}$, $Z_{i1}$, etc.

The $\Delta Y(X)$ method uses ``wide" data.
That is, the observed data have one outcome \emph{change} observation for each unit and post-treatment period (relative to a pre-treatment reference period).
For instance, given pre-treatment period $t=0$, we have $\Delta Y_{it} = Y_{it} - Y_{i0}$, for $t \in \{1, 2\}$.  
We regress these outcome changes on an indicator of treatment group and covariates,
\begin{equation} \label{eq:est_deltaY}
    \Delta Y_{it} \sim \theta A_{i1} + \bm \beta' \mathbf{X}_{i}\;, 
\end{equation}
taking the coefficient $\theta$ as the treatment effect estimate.

The TWFE estimator, $Y(X)$, uses ``long'' data in which we have outcome observations for each unit in each time period.
We fit a regression model to the longitudinal outcomes, 
\begin{equation} \label{eq:est_twfe}
    Y_{it} \sim  \eta_0 + \eta_A A_{i1} + \eta_P P_t + \theta A_{i1} P_t + \bm{\beta}'_t \mathbf{X}_{i}\;,
\end{equation}
where $P_t = I\{t\geq t_0\}$ for $t_0$ an indicator of the first post-treatment period.
The estimated coefficient $\theta$ on the interaction term is taken as an estimate the treatment effect estimate.
Note that we have given the covariates time-varying coefficients ${\bm \beta}_t$.

To implement the de Chaisemartin and D'Haultfœuille's estimator,  $\epsilon(\Delta Y(\Delta X))$,\cite{dechaisemartinTwoWayFixedEffects2020}, we use the \texttt{did\_multiplegt} function from the \texttt{DIDmultiplegt} package with the \texttt{mode="old"} option.
This function operates on covariate changes, $\Delta \mathbf{X}_{it} = \mathbf{X}_{it}-\mathbf{X}_{it_0}$, where we have extended out notation slightly to indicate that changes at each time point are taken relative to a reference period $t_0$.
Then the function regresses outcome changes on covariate changes in the comparison group only,
$$
\Delta Y_{it} \sim \bm{\beta}'\Delta \mathbf{X}_{it} \; | \; A_{i1}=0\;.
$$
From this, we obtain predicted outcome changes for each unit via $\widehat{\Delta Y}_{it}=\hat{\bm\beta}'\Delta \mathbf{X}_{it}$,
and residuals, $\epsilon_{it} = \Delta Y_{it} - \widehat{\Delta Y}_{it}$.
Finally, we take the difference in these residuals between treated and control units as the estimate of the treatment effect,
\begin{equation}\label{eq:est_dh}
    \mathbb{E}_{trt}(\epsilon_{it}) - \mathbb{E}_{ctrl}(\epsilon_{it})\;,
\end{equation}
where for a generic $x_i$, $\mathbb{E}_{trt}(x_i)=\frac{1}{|\mathcal{N}^{trt}|}\sum_{i \in \mathcal{N}^{trt}} x_i$.

We fit Heckman's outcome regression, Abadie's IPW, and Sant'Anna and Zhao's DR estimators using  the \texttt{did} package \cite{callawayDidTreatmentEffects2022} via the \texttt{att\_gt} function.
For the first of these, Heckman's outcome regression, $\epsilon(\Delta Y(X)$, (using option \texttt{est\_method="reg"}) we regress outcome changes on covariates in the comparison group only,
\begin{equation} \label{eq:est_cs_m}
     \Delta Y_{it} \sim  \bm{\beta}' \mathbf{X}_i \; | A_{i1} =0\;,
\end{equation}
which yields predicted outcome changes for treated units via $\widehat{\Delta Y}_{it} = \hat{\bm \beta}'\mathbf{X}_i$.
The estimator is simply the difference between observed and predicted outcome changes in the treated group,
\begin{equation}\label{eq:est_heckman}
\mathbb{E}_{trt}\left(\Delta Y_{it} - \widehat{\Delta Y}_{it} \right)\;.
\end{equation}

Abadie's IPW estimator, $w(X)\Delta Y$, (fit using option \texttt{est\_method="ipw"}) omits outcome regression and instead adjusts for covariates using propensity scores. 
In the propensity score model, we regress the treatment indicator on the covariates,
\begin{equation} \label{eq:est_ps_model_static}
    Pr(A_{i1} = 1) \sim \mbox{logit}^{-1}(\bm{\pi}' \mathbf{X}_i)\;,
\end{equation}
which yields propensity scores for any unit via $\hat{e}\left(\mathbf{X}_i\right) = \mbox{logit}^{-1}\left(\hat{\bm \pi}'\mathbf{X}_i\right)$.
These are converted into weights for use in weighted regression.
The weights for treated units are $w_i \propto 1$, while the weights for control units are 
$$
w_i \propto \frac{\hat{e}\left(\mathbf{X}_{i}\right)}{1 - \hat{e}\left(\mathbf{X}_{i}\right)}\;.
$$
Normalizing constants ensure that each set of weights sum to one.
The estimator is the weighted difference in outcome changes, 
\begin{equation}\label{eq:est_ipw}
\mathbb{E}_{trt}\left(w_i \Delta Y_{it} \right)-
\mathbb{E}_{ctrl}\left(w_i \Delta Y_{it} \right)\;.
\end{equation}

Finally, the doubly robust estimator, $w(X)\Delta Y$, fit using (option \texttt{est\_method="dr"}) combines the output of both models as
\begin{equation} \label{eq:est_cs_dr}
\mathbb{E}_{trt}\left[\hat{w}_i\left(\Delta Y_i - \widehat{\Delta Y}_i\right)\right] - 
    \mathbb{E}_{ctrl}\left[\hat{w}_i\left(\Delta Y_i - \widehat{\Delta Y}_i\right)\right]\;,
\end{equation}

We consider two propensity score weighting procedures from Stuart et al.~\cite{stuartUsingPropensityScores2014a}.
In the first approach, $w_g(X)Y$, (the authors' preferred) we weight three of the groups in the 2x2 DiD table toward a single reference group.
We first construct a four-level indicator variable in our ``long'' data, where 
$g=1$ is treated, pre (the reference group); $g=2$ is comparison, pre; $g=3$  is comparison, post; and $g=4$ is treated, post.
Then we fit a log-linear multinomial model to the group indicator,
\begin{equation}\label{eq:est_ps_4grp}
Pr(G_i=g) = \exp\left(\mathbf{\bm\pi}_g'\mathbf{X}_i\right)\;
\end{equation}
using the \texttt{multinom} function of the \texttt{nnet} package \cite{venablesModernAppliedStatistics2002}, which fits these models using neural nets.
Then we use the model's predicted probabilities of membership in each group to construct weights for each unit,
$$
w_i = \frac{\hat{e}_1(\mathbf{X_{it}})}{\hat{e}_{G_i}\left(\mathbf{X}_{it}\right)}\;,
$$
where $\hat{e}_g(\mathbf{X}_i)$ is the model-predicted probability of belonging to group $g$ and $G_i$ is the index of the group to which the unit belongs.

The other estimator, $w_t(X)Y$, fits separate propensity score models for the pre and post periods using long data and two logistic models,
\begin{equation}\label{eq:est_ps_2time}
Pr(A_{i1}=1|\mathbf{X}_{it}) = \mbox{logit}^{-1}\left(\bm\pi_t'\mathbf{X}_{it}\right) \;,
\end{equation}
which yield propensity scores for each unit via $\hat{e}_t(\mathbf{X}_{it}) = \mbox{logit}^{-1}\left(\hat{\bm \pi}'_t\mathbf{X}_{it}\right)$.
From these, we can construct the following weights
\begin{equation}\label{est:stuart_ate_wts}
w_{it} =
\left\{
\begin{array}{cl}
\frac{1}{\hat{e}_t\left(\mathbf{X}_{it}\right)} & ,\; A_{i1} = 1, \\
\frac{1}{1 - \hat{e}_t\left(\mathbf{X}_{it}\right)} & ,\; A_{i1} = 0\;.
\end{array}
\right.
\end{equation}

For both estimators, we simply fit a weighted linear regression with the weights given above,
\begin{equation}\label{eq:est_stuart_reg}
    Y_{it} \sim \eta_0 + \eta_{A} A_{i1} + \eta_{P} P_t + \theta A_{i1} P_t\;,
\end{equation}
and take the coefficient $\theta$ as the treatment effect estimate.

Finally, we consider two hybrid estimation procedures from Myint~\cite{myintControllingTimevaryingConfounding2023}. 
Myint compared sequential exchangeability and parallel trends identification strategies, but we focus on the estimators of group-time ATT's, of which there are 6 in the Myint paper.
Two combine time-varying treatment weights with marginal structural models for estimation, while two use either ATT or ATE weights with the C-S'A estimator.

The propensity score model regresses the treatment indicator on covariates as in Eq.~\ref{eq:est_ps_model_static}, which produces the usual estimated propensity scores.
From these, we form stabilized ATE weights
\begin{equation}\label{eq:myint_ATE_wts}
w_{it} =
\left\{
\begin{array}{cl}
 \frac{p^{trt}}{\hat{e}\left(\mathbf{X}_{it}\right)} & ,\; A_{i1} = 1 \\
 \frac{1-p^{trt}}{1 - \hat{e}\left(\mathbf{X}_{it}\right)} & ,\; A_{i1} = 0\;,
\end{array}
\right.
\end{equation}
and ATT weights
\begin{equation}\label{eq:myint_ATT_wts}
w_{i} =
\left\{
\begin{array}{cl}
 1 & ,\; A_{i1} = 1 \\
 \frac{\hat{e}\left(\mathbf{X}_{it}\right)}{1 - \hat{e}\left(\mathbf{X}_{it}\right)} & ,\; A_{i1} = 0\;.
\end{array}
\right.
\end{equation}

Finally, the estimator is a weighted difference in outcome changes between treated and control,
\begin{equation}\label{eq:myint}
    \mathbb{E}_{trt}\left(w_{it} \Delta Y_{it} \right)-
    \mathbb{E}_{ctrl}\left(w_{it} \Delta Y_{it} \right)\;.
\end{equation}
and take the coefficient $\theta$ as the treatment effect.

\end{document}

%% file: preamble.tex
\usepackage{acronym}
\usepackage{graphicx}
\usepackage[margin=2.5cm]{geometry}
\usepackage{setspace}
\usepackage{url}
\usepackage[square,numbers,sort&compress]{natbib}

\usepackage{multirow}
\usepackage{tikz}
\usetikzlibrary{arrows, fit, arrows.meta}

\usepackage{tcolorbox}
\usepackage{ragged2e}
\usepackage{enumitem}
\usepackage[dvipsnames,table,xcdraw]{xcolor}
\usepackage{subcaption}

\usepackage{array}

\usepackage{pdflscape}
\usepackage{xltabular}
\newcolumntype{L}{>{\RaggedRight\hangafter1\hangindent1em}X}
\newcolumntype{O}{>{\RaggedRight\hspace{0pt}}X}
\usepackage{booktabs}
\newlength\mylen
\usepackage[labelfont=bf,skip=5pt]{caption} 

\usepackage{booktabs}
\usepackage{multirow}
\usepackage[colorlinks=true, linkcolor=black, citecolor=black, urlcolor=black]{hyperref}

\acrodefplural{DAG}[DAGs]{Directed Acyclic Graphs}
\acrodefplural{SCM}[SCMs]{Structural Causal Models}
\acrodefplural{SWIG}[SWIGs]{Single World Intervention Graph}

\acrodef{DiD}{Difference-in-Differences}
\acrodef{ATT}{Average Treatment effect on the Treated}
\acrodef{ATE}{Average Treatment Effect}
\acrodef{CATT}{Conditional ATT}
\acrodef{IPW}{Inverse Probability Weighting}
\acrodef{DAG}{Directed Acyclic Graph}
\acrodef{TWFE}{Two-Way Fixed Effects}
\acrodef{CPC+}{Comprehensive Primary Care Plus}
\acrodef{SCM}{Structural Causal Model}
\acrodef{IV}{Instrumental Variable}
\acrodef{OLS}{Ordinary Least Squares}
\acrodef{SWIG}{Single World Intervention Graph}

\usepackage{amsmath,amssymb,bm}

\usepackage{authblk}

\usepackage{float}

\title{\LARGE\textbf{A formal approach to variable selection in difference-in-differences}\thanks{We thank Katherine Ianni, Alyssa Chen, Chi Zhang, Karim Chalak, Carlos Cinelli, Elias Bareinboim, Nima Hejazi and participants of the 2025 Society of Epidemiology Research (SER) conference, the 2025 International Conference on Health Policy Statistics (ICHPS) and the Health Economics seminar at Harvard's Department of Health Care Policy for helpful discussions. This project was funded under grant number R01HS028985 from the Agency for Healthcare Research and Quality (AHRQ), U.S. Department of Health and Human Services (HHS). The authors are solely responsible for this document’s contents, findings, and conclusions, which do not necessarily represent the views of AHRQ.}}

\author[1]{Daniela Rodrigues}
\author[2]{Laura A. Hatfield}

\affil[1]{Department of Health Care Policy, Harvard Medical School, Boston, MA, USA}
\affil[2]{Statistics and Data Science Department, NORC at the University of Chicago, Chicago, IL, USA}

\date{\today}